\documentclass[conference]{IEEEtran}
\IEEEoverridecommandlockouts

\usepackage{cite}
\usepackage{amsmath,amssymb,amsfonts}
\usepackage{algorithmic}
\usepackage{graphicx}
\usepackage{textcomp}
\usepackage{xcolor}
\usepackage[hyphens]{url}
\usepackage{minted}
\usepackage{listings}
\usepackage{xcolor}

\usepackage{fancyhdr}
\usepackage{hyperref}
\usepackage{pifont}
\usepackage{multirow}
\definecolor{ForestGreen}{RGB}{34, 139, 34} 
\usepackage{algorithm}
\usepackage{booktabs}
\usepackage{caption} 

\def\BibTeX{{\rm B\kern-.05em{\sc i\kern-.025em b}\kern-.08em
    T\kern-.1667em\lower.7ex\hbox{E}\kern-.125emX}}
\begin{document}

\pdfpagewidth=8.5in
\pdfpageheight=11in

\newcommand{\iscasubmissionnumber}{NaN}

\pagenumbering{arabic}

\title{CODO: An Automated Compiler for Comprehensive \\
Dataflow Optimization}
\author{\normalsize{ISCA 2026 Submission
    \textbf{\#88} -- Confidential Draft -- Do NOT Distribute!!}}







\author{%
\begin{tabular}{@{}c@{\hspace{0.6cm}}c@{\hspace{0.6cm}}c@{}}

\begin{tabular}[t]{@{}c@{}}
{ Weichuang Zhang}\\
School of Computer Science\\
Shanghai Jiao Tong University\\
Shanghai, China\\
1064080006@sjtu.edu.cn
\end{tabular}
&
\begin{tabular}[t]{@{}c@{}}
{ Yiquan Wang}\\
School of Computer Science\\
Shanghai Jiao Tong University\\
Shanghai, China\\
abcdfehg@sjtu.edu.cn
\end{tabular}
&
\begin{tabular}[t]{@{}c@{}}
{ Xinzhou Zhang}\\
School of Computer Science\\
Shanghai Jiao Tong University\\
Shanghai, China\\
xz\_zhang@sjtu.edu.cn
\end{tabular}

\\[6em]

\begin{tabular}[t]{@{}c@{}}
{ Chi Zhang}\\
School of Computer Science\\
Shanghai Jiao Tong University\\
Shanghai, China\\
zhang-chi@sjtu.edu.cn
\end{tabular}
&
\begin{tabular}[t]{@{}c@{}}
{ Yu Feng}\\
School of Computer Science\\
Shanghai Jiao Tong University\\
Shanghai, China\\
y-feng@sjtu.edu.cn
\end{tabular}
&
\begin{tabular}[t]{@{}c@{}}
{ Xiaofeng Hou}\\
School of Computer Science\\
Shanghai Jiao Tong University\\
Shanghai, China\\
hou-xf@cs.sjtu.edu.cn
\end{tabular}

\\[6em]

\begin{tabular}[t]{@{}c@{}}
{ Chao Li}\\
School of Computer Science\\
Shanghai Jiao Tong University\\
Shanghai, China\\
lichao@cs.sjtu.edu.cn
\end{tabular}
&
\begin{tabular}[t]{@{}c@{}}
{ Jieru Zhao\IEEEauthorrefmark{1}}\\
School of Computer Science\\
Shanghai Jiao Tong University\\
Shanghai, China\\
zhao-jieru@sjtu.edu.cn
\end{tabular}
&
\begin{tabular}[t]{@{}c@{}}
{ Minyi Guo}\\
Guizhou Provincial Laboratory of Big Data \\
College of Computer Science \\
 and Technology, Guizhou University \\
School of Computer Science, SJTU\\
guo-my@cs.sjtu.edu.cn
\end{tabular}

\end{tabular}%
\thanks{\IEEEauthorrefmark{1} Corresponding author: Jieru Zhao.}%
}


\maketitle
\thispagestyle{plain}
\pagestyle{plain}



\begin{abstract}

FPGAs are well-suited for dataflow architectures that process data in a streaming or pipelined manner, thus satisfying the high computational and communication demands of emerging applications. However, manually implementing an efficient dataflow architecture for large-scale applications is still challenging, even for specialists who use high-level synthesis (HLS) to simplify FPGA programming. 

To address this, we introduce CODO, an automated compiler that generates feasible and efficient dataflow accelerators on FPGAs. CODO features a systematic method for detecting and eliminating both coarse-grained and fine-grained dataflow violations. Building on this, CODO performs both on- and off-chip data movement optimizations to maximize transfer efficiency.
To guarantee a higher design quality, CODO performs automatic scheduling to generate high-performance dataflow accelerators, ensuring a balanced performance-resource trade-off.  
Synthesis results show that CODO delivers $1.45\times$
to $4.52\times$ latency speedups on typical computation kernels and 
$3.7\times$ to $33.8\times$ speedups on DNN models compared to SOTA frameworks. 
In on-board evaluations, CODO achieves $7.3\times$ average speedup on CNN models and $2.07\times$ average speedup on the GPT-2 model over SOTA frameworks. 
The compiler is open-sourced at \url{https://github.com/sjtu-zhao-lab/codo-artifact}.

\end{abstract}
\section{Introduction}
\vspace{-0.1cm}

Dataflow architectures are suitable for workloads that require massive data movement and operations due to their low latency
\cite{abts2020think,jia2021tensorlib,gobieski2022riptide}. 
Fundamentally, these architectures leverage task-level pipelining to allow distinct functions and loops to overlap in their execution, rather than running sequentially. Furthermore, they exploit efficient on-chip communication between tasks \cite{ye2024hida,streamtensorMICRO25}, significantly reducing the overhead of frequent external memory accesses \cite{chenTRETS2024,nowatzki2017ISCA}. 
FPGAs, with their reconfigurable logic and customizable data paths, are well-suited for implementing dataflow accelerators that process data in a streaming or pipelined manner \cite{zhang2018dnnbuilder,kwon2021heterogeneous,lian2019high,andri2017yodann}. 
Note that while the term \textit{dataflow} is also used to describe dynamic scheduling-based \textit{dataflow} circuits \cite{dynamicHLS2018} or \textit{dataflow} mapping strategies like input/output stationary \cite{datareuse2017}, these concepts are conceptually orthogonal to the dataflow architecture discussed in this paper.

However, the high efficiency of dataflow accelerators comes at the expense of a complex design process using hardware description languages (HDLs). To simplify FPGA development, developers utilize high-level synthesis (HLS) to translate C/C++ code into HDL implementations automatically \cite{jasoncong2011hls}. 
Nevertheless, a notable gap still persists between HLS programming and efficient dataflow implementations. 
Commercial HLS tools, such as AMD Vitis HLS \cite{vitishls}, provide basic dataflow scheduling primitives, i.e., \textit{the dataflow pragma} \cite{hlsdataflow}, to enable pipelined execution between loops or functions. However, this optimization works only if the coding styles satisfy stringent requirements \cite{hlsdataflow}, which many handcrafted algorithms fail to meet. This mismatch hinders effective optimization or even leads to synthesis failures.
Consequently, developers must perform extensive code refactoring and optimization manually to convert algorithms into dataflow-feasible formats and produce dataflow accelerators.

\textbf{Prior Research.} To reduce developing efforts, prior methods 
enhance programming efficiency using domain-specific languages (DSLs) \cite{lai2019heterocl, xiang2022heteroflow, zhang2024pom}, or directly parse C++ inputs or PyTorch models into intermediate representations (IRs) \cite{ye2022scalehls, huang2021pylog, 2022fplpolsca}.
They primarily focus on kernel computation optimization, with limited consideration for dataflow optimization. Recently, several compilers have been proposed for dataflow optimization across multiple kernels or tasks \cite{ye2024hida, chen2024allo, streamhlsFPGA25, streamtensorMICRO25}, enabling automatic generation of dataflow accelerators.
However, these approaches fail to fully resolve potential issues (Fig. \ref{fig:motivating-exp}) in the dataflow,
limiting their ability to further optimize code and exploit parallelism. As a result, the generated designs may suffer from suboptimal performance or even encounter deadlocks when deployed on FPGA boards. 

\textbf{Key Idea.} 
The performance of dataflow accelerators is influenced by multiple factors. Fundamentally, the input code must satisfy strict constraints to enable correct streaming execution (\textit{correctness}). On top of that, achieving high-throughput communication necessitates effective optimization of communication buffers and careful alignment of computation patterns between adjacent tasks to enhance data transfer efficiency (\textit{communication}). 
Finally, balancing task latencies through techniques such as loop tiling, unrolling, and pipelining is essential for improving overall performance (\textit{parallelism}). 

The core problem is that these factors are deeply codependent, yet prior work typically handles them in a decoupled manner. This leads to conflicts where optimizing one aspect in isolation negatively impacts the others. For instance, aggressive code transformations to meet dataflow constraints may result in inefficient computation and memory access patterns that create a communication bottleneck. Conversely, communication or parallelism optimizations may violate dataflow constraints and produce invalid designs.
Moreover, the growing scale and structural complexity of modern DNNs further exacerbate the problem, making it increasingly difficult to construct deep, high-throughput pipelines for large models. 

To overcome these issues, we build a compiler that jointly co-optimizes correctness, communication, and parallelism. Through advanced code analysis, versatile optimization techniques, and automated scheduling, the compiler performs coordinated transformations that harmoniously benefit all three aspects rather than creating conflicts, automatically generating high-performance accelerators for large-scale models.




\textbf{Challenges.} Achieving the goal is challenging. Firstly, the input algorithm may violate dataflow constraints, resulting in \textit{dataflow violations} that must be eliminated. At the coarse-grained level, existing HLS tools impose a strict single-producer-consumer constraint to enable dataflow optimization \cite{hlsdataflow}. At the fine-grained level, producers and consumers must maintain consistent data access order and count to ensure correct and efficient streaming execution.
Although recent works attempt to address these violations \cite{ye2024hida, chen2024allo, streamhlsFPGA25, streamtensorMICRO25}, their methods are difficult to fully eliminate all violations in large-scale models. Consequently, unresolved violations result in \textit{discontinuous dataflow regions}, breaking end-to-end streaming (Fig. \ref{fig:motivating-exp}, Issue 1).

\begin{figure}
    \centering
    \includegraphics[width=0.4\textwidth]{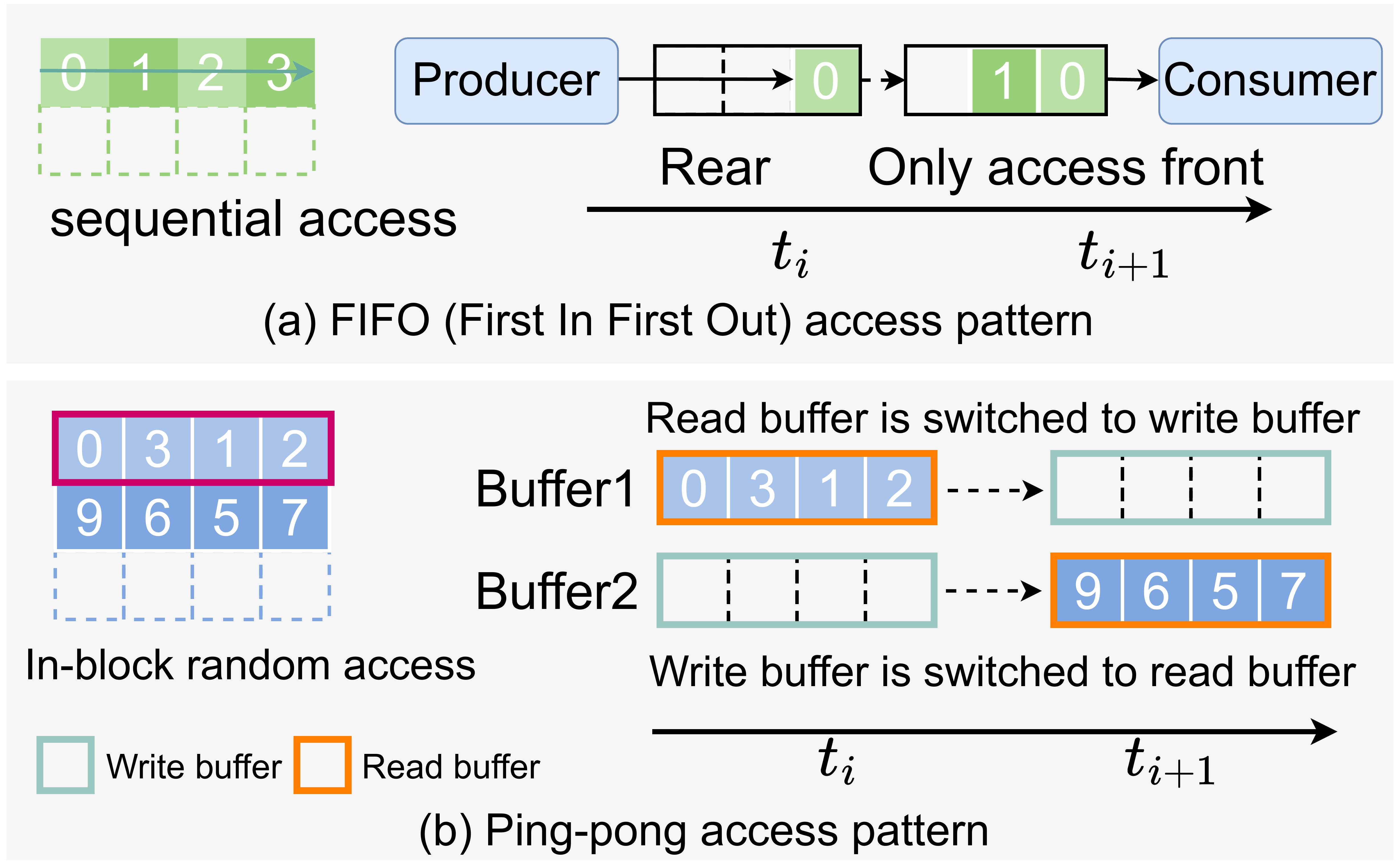}
    \caption{Dataflow execution with FIFO and ping-pong buffer. Numbers in (a) and (b) represent the data access order.}
    \label{fig:fifo}
    \vspace{-0.4cm}
\end{figure}

Secondly, data is transferred through communication buffers, typically implemented as ping-pong buffers or FIFOs (First-In-First-Out), as shown in Fig. \ref{fig:fifo}.
To achieve high throughput, intermediate results must be produced and consumed just-in-time to prevent pipeline stalls. 
This requires careful selection of buffer types.
Moreover, this idealized communication is often disrupted by other optimizations, such as violation elimination, which may unintentionally alter computation schedules and compromise communication efficiency. Without holistic coordination and advanced code analysis, such issues are easily overlooked \cite{streamhlsFPGA25}, resulting in \textit{delayed buffer writes and performance degradation} (Fig.~\ref{fig:motivating-exp}, Issue 2).

Thirdly, improving dataflow performance requires balancing tasks through techniques such as loop tiling, unrolling, and pipelining. This becomes even more challenging in FIFO-based dataflow, as code optimizations affect data access patterns, requiring careful coordination between adjacent
producers and consumers to avoid new dataflow violations. Consequently, existing auto-scheduling methods for ping-pong-based dataflow \cite{ye2024hida} are not directly applicable. While some tools use manual scheduling \cite{lai2019heterocl, xiang2022heteroflow} or nonlinear programming (NLP)-based methods \cite{streamhlsFPGA25, unifiedFPGA2025} for FIFO-based dataflow, they lack effective pruning methods to handle the exponentially increasing design space. As a result, these approaches \textit{fail to scale to large-scale models}.


\textbf{Our Solution.} To tackle the above challenges, we propose CODO, an open-source compiler that performs comprehensive dataflow optimizations and automatically generates high-performance accelerators. CODO resolves dataflow violations, ensures communication efficiency, and explores resource-aware parallelism strategies to guarantee balanced task execution for large-scale models.

\begin{table*}[]
\renewcommand\arraystretch{1}
\caption{Comparison between representative compilers.}
\vspace{-0.2cm}
\label{comparison-table}
\resizebox{\textwidth}{!}
{
\begin{tabular}{lccccccc}
\hline
\textbf{Feature}     & \textbf{ScaleHLS \cite{ye2022scalehls}}   &  \textbf{POM \cite{zhang2024pom}} & \textbf{Allo \cite{chen2024allo}} & \textbf{HIDA \cite{ye2024hida}} & \textbf{StreamHLS \cite{streamhlsFPGA25}} & \textbf{StreamTensor \cite{streamtensorMICRO25}} & \textbf{CODO  }         \\ \hline
Compiler Front-end  &  PyTorch &  DSL &  PyTorch & DSL  & PyTorch & PyTorch & PyTorch     \\ 

Coarse-grained Violation Elimination  & \color{orange}{Limited}  & \color{orange}{Manual}   & \color{orange}{Manual}  & \color{ForestGreen}{\ding{52}}  & \color{ForestGreen}{\ding{52}}  & \color{ForestGreen}{\ding{52}} & \color{ForestGreen}{\ding{52}}      \\ 

Fine-grained Violation Elimination  & \color{red}{\ding{56}}  & \color{red}{\ding{56}}   & \color{red}{\ding{56}}  & \color{red}{\ding{56}} & \color{orange}{Limited}  & \color{orange}{Limited}  & \color{ForestGreen}{\ding{52}}         \\ 

Efficient Communication Buffer & \color{red}{\ding{56}}  & \color{red}{\ding{56}}   & \color{ForestGreen}{\ding{52}} & \color{ForestGreen}{\ding{52}} & \color{red}{\ding{56}}  & \color{red}{\ding{56}}  & \color{ForestGreen}{\ding{52}}         \\ 

Resource-aware Node Balancing   & \color{red}{\ding{56}}   & \color{red}{\ding{56}}  & \color{red}{\ding{56}}   & \color{red}{\ding{56}} & \color{red}{\ding{56}} & \color{red}{\ding{56}} & \color{ForestGreen}{\ding{52}}         \\ 


Automated Scheduling or DSE     & \color{ForestGreen}{\ding{52}}     & \color{ForestGreen}{\ding{52}}  & \color{red}{\ding{56}}    & \color{ForestGreen}{\ding{52}}   & \color{ForestGreen}{\ding{52}} & \color{ForestGreen}{\ding{52}} & \color{ForestGreen}{\ding{52}}  \\

On-board Verification  & \color{red}{\ding{56}}  & \color{red}{\ding{56}}  & \color{ForestGreen}{\ding{52}}  & \color{red}{\ding{56}} & \color{red}{\ding{56}} & \color{ForestGreen}{\ding{52}} & \color{ForestGreen}{\ding{52}}       \\ 

Open Source Project   & \color{ForestGreen}{\ding{52}}     & \color{ForestGreen}{\ding{52}}   & \color{ForestGreen}{\ding{52}}   & \color{ForestGreen}{\ding{52}} & \color{ForestGreen}{\ding{52}} & \color{red}{\ding{56}} & \color{ForestGreen}{\ding{52}}  \\

\hline
\end{tabular}%
}
\vspace{-0.4cm}
\end{table*}
\begin{itemize}
    \item We present an end-to-end compiler that automatically transforms an input algorithm into high-quality dataflow accelerators, along with the host code.
    \item We eliminate both coarse- and fine-grained dataflow violations and enable efficient data communication via on- and off-chip optimizations.
    \item We propose an automated scheduling method that rapidly determines suitable parallelism strategies with high resource efficiency to generate a high-performance design.
    \item We perform synthesis and on-board evaluations. CODO achieves $3.7\times$ to $33.8\times$ speedup on DNN models in synthesis and an average speedup of $7.3\times$ for DNNs and $2.07\times$ for GPT-2 on-board compared to SOTA compilers.
\end{itemize}

\vspace{-0.1cm}
\section{Background and Motivation}\label{Background and Motivation}
\vspace{-0.1cm}

\subsection{Dataflow Architecture and its Violations} \label{sec:background-violation}
A dataflow architecture is a computing model where operations are triggered by the availability of input data, enabling efficient parallel execution. 

\textbf{FIFO vs. Ping-Pong Buffer.} A critical aspect of dataflow architecture is the way of data processing and transmission between tasks. Two common patterns are  shown in Fig. \ref{fig:fifo}. 

For \textbf{FIFO-based dataflow}, data elements are generated sequentially by the producer and streamed into a first-in, first-out buffer. The consumer then reads and processes the data one element at a time following their order of arrival. FIFO-based dataflow accelerators often achieve higher performance, since consumers can start execution immediately once the required data element is available. Resource utilization is also efficient, since only the necessary in-flight data needs to be stored. 

In contrast, \textbf{ping-pong-based dataflow} groups data elements into blocks. Adjacent blocks are written alternately into two separate buffers, e.g., Buffer1 and Buffer2 in Fig.~\ref{fig:fifo}(b), leading to higher memory usage. While the producer writes to one buffer, the consumer reads from the other. Unlike FIFO, ping-pong-based dataflow inherently exhibits higher latency, as the consumer cannot begin execution until the entire data block is available. 
Nevertheless, data in each block can be accessed randomly, offering greater flexibility in data processing.

Commercial HLS tools enable dataflow processing on FPGAs using certain directives or pragmas, such as \textit{\#pragma HLS dataflow} in Vitis HLS. However, mapping computation graphs directly 
can lead to dataflow violations, which can be classified into \textit{coarse-grained} and \textit{fine-grained} categories.

\textbf{Coarse-grained Dataflow Violation.}
Existing HLS tools enforce a strict single-producer, single-consumer constraint, requiring data to flow exclusively from one producer to one consumer throughout the design \cite{hlsdataflow}. However, the topologies, data dependencies, and the inherent dataflow in real-world applications, such as DNN models, are complex, frequently violating this strict constraint and hindering the task-level parallelism. We term this violation as \textit{coarse-grained dataflow violation}. 
While tools like Vitis HLS offer guidelines for handling such violations, developers must still manually refactor the code, making this process labor-intensive and error-prone, especially for complex applications. 

\textbf{Fine-grained Dataflow Violation.}
Existing frameworks \cite{ye2022scalehls,zhang2024pom,ye2024hida} prefer ping-pong-based dataflow due to its flexibility in data access order, albeit with some performance loss. In contrast, FIFO-based dataflow can deliver higher performance but is more challenging to implement, as the strict sequential access constraint of FIFO demands fine-grained computation control and significant code refactoring. We term the corresponding violation as \textit{fine-grained dataflow violation}.
Specifically, mismatched \textbf{data access count} or \textbf{order} between producers and consumers can cause issues such as data loss, FIFO overflows or underflows, and even deadlocks. Worse still, these violations may not be detected during synthesis. Although HLS tools offer co-simulation to identify potential deadlocks, it can take days for large models, and even fail to detect issues reliably. 
Therefore, detecting and eliminating these violations at an early stage is
essential yet challenging.

\begin{figure*}[h]
    \centering
    \includegraphics[scale=0.888]{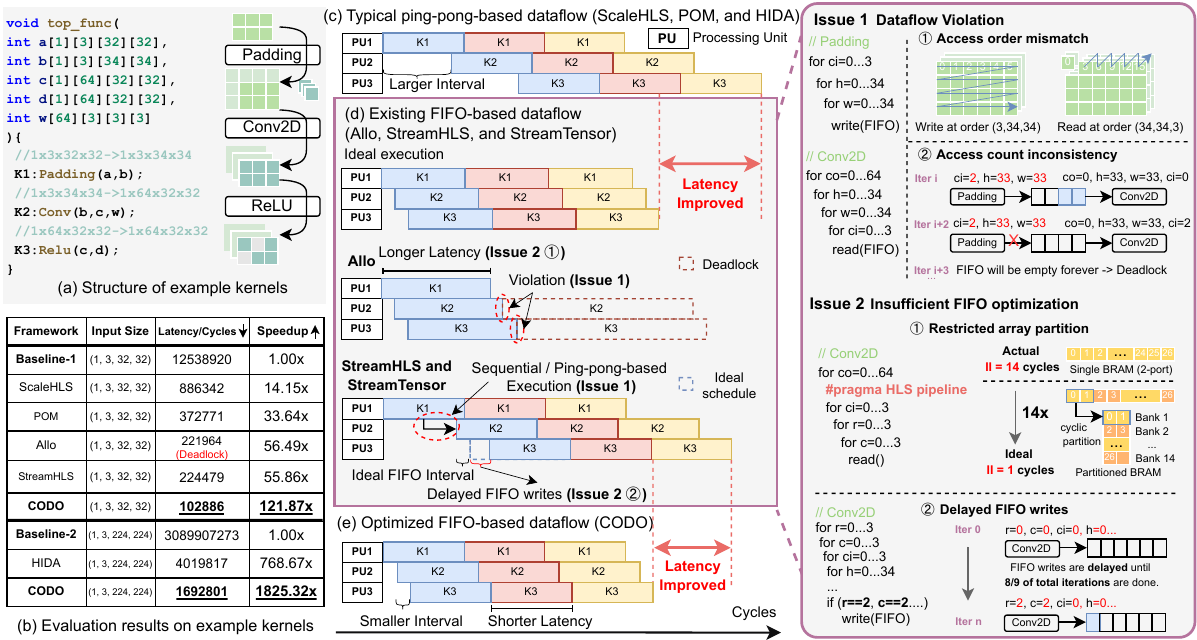}
    \caption{Motivating example. (a) The code snippet consists of a top function and three sub-functions: Padding, Convolution, and ReLU.
    (c)(d)(e) present the schedules adopted by different frameworks, including ScaleHLS \cite{ye2022scalehls}, POM \cite{zhang2024pom},  HIDA \cite{ye2024hida}, Allo \cite{chen2024allo}, StreamHLS \cite{streamhlsFPGA25}, StreamTensor \cite{streamtensorMICRO25}, and CODO. K1, K2, K3 correspond to example kernels.
    Different colors distinguish different iterations. 
    (b) shows the performance of code without optimization (baseline) as well as the optimized code generated by six open-source compilers. The target device is AMD Alveo U280 FPGA \cite{alveou280}. The performances are \textit{HLS synthesis results}.}

    \label{fig:motivating-exp}
    \vspace{-0.4cm}
\end{figure*}

\vspace{-0.1cm}
\subsection{Related Work and Comparison}\label{related work}
We compare representative FPGA compilers 
in Table \ref{comparison-table}. 
Recent achievements in domain-specific languages (DSLs) for FPGA enhance productivity \cite{spatial2018, aetherling2020, lai2019heterocl, xiang2022heteroflow, huang2021pylog, zhang2024pom}. Spatial \cite{spatial2018} and Aetherling \cite{aetherling2020} provide specific coding styles or rewriting rules to generate high-performance HDL code in Chisel \cite{chisel2012}. 
Dahlia \cite{dahlia2020pldi} introduces an HLS language with a type system that explicitly tracks hardware resources and time ordering when resources are available.
HeteroCL \cite{lai2019heterocl} and HeteroFlow \cite{xiang2022heteroflow} extend the TVM DSL, providing customization for computation, memory, and data movement. 
POM \cite{zhang2024pom} delivers performance improvements
through loop transformations and dependency analysis.  
Besides DSL, some approaches directly parse C++ or Python inputs into IRs or provide  templates to generate efficient FPGA accelerators. 
PyLog \cite{huang2021pylog} can process Python code and offer automatic pragma insertion.
Sisyphus \cite{Sisyphusfpga25} proposes an optimization template that integrates code transformation, pragma insertion, and tile size selection into a single optimization problem. Prometheus \cite{Prometheus2025} further enhances Sisyphus by providing unified optimization of computation and communication. ScaleHLS \cite{ye2022scalehls} parses PyTorch models into IRs based on MLIR for further optimization.

Despite programming efficiency and performance improvement, these frameworks primarily focus on optimizing kernel computation,
with limited consideration for dataflow optimization of large-scale applications.
For frameworks enabling automated scheduling or design space exploration (DSE) \cite{zhang2024pom, ye2022scalehls}, as shown in Table \ref{comparison-table}, 
ScaleHLS partially mitigates coarse-grained dataflow violation by resolving multi-consumer violations for ping-pong-based dataflow, whereas POM relies on developers to manually manage data dependencies using DSL to prevent violations. However, both methods apply homogeneous optimization strategies across different tasks, e.g., layers of DNN models, resulting in suboptimal designs.

To boost performance, compilers for dataflow optimization are proposed \cite{guo2023tapa,ye2024hida, chen2024allo, streamhlsFPGA25, streamtensorMICRO25}.
TAPA \cite{guo2023tapa} provides specialized APIs to express dataflow. However, it requires manual scheduling and lacks in-depth exploration of parallelism. 
Allo \cite{chen2024allo} features a composable DSL for efficient spatial accelerator generation with FIFOs.
However, it overlooks dataflow violations in the design, potentially incurring pipeline stalls or even deadlocks.
HIDA \cite{ye2024hida} targets ping-pong-based dataflow accelerators, eliminates coarse-grained dataflow violations, and applies dataflow parallelization automatically, but its processing in data blocks still results in suboptimal performance.
StreamHLS \cite{streamhlsFPGA25} partially resolves dataflow violations through reordering loops and adding control statements for a single task. However, lacking advanced code analysis and scalable optimization methods, they fail to fully address violations for large-scale models featuring complex code patterns and deep layer hierarchies. 
StreamTensor \cite{streamtensorMICRO25} introduces an iterative tensor-based type system for dataflow optimization, but it falls back to a conservative ping-pong buffering strategy when encountering fine-grained dataflow violations, limiting its ability to exploit high-performance FIFO-based designs.

In contrast, CODO enables efficient streaming execution by eliminating dataflow violations at both coarse-grained and fine-grained levels through advanced code analysis and transformation, applying extensive memory optimization for efficient data communication, and providing automated resource-aware scheduling to reduce overall latency with minimal resource overhead. CODO automatically generates efficient dataflow accelerators with host code for control, which can be directly deployed on an FPGA board. 



\subsection{Motivating Example} \label{sec:moti}
To further illustrate differences, we use a motivating example with three tasks (kernels), as shown in Figure \ref{fig:motivating-exp}(a).
The dataflow patterns in existing works can be categorized into ping-pong-based and FIFO-based approaches.

\textbf{Typical ping-pong-based Approaches.} As shown in Fig. \ref{fig:motivating-exp}(c), POM \cite{zhang2024pom}, ScaleHLS \cite{ye2022scalehls}, and HIDA \cite{ye2024hida} typically employ a ping-pong-based dataflow schedule. 
This approach requires waiting for an entire data block to be produced before subsequent kernels can start execution, resulting in long intervals and limited opportunities for overlapping operations.
In contrast, as illustrated in Fig. \ref{fig:motivating-exp}(d), an ideal FIFO-based execution allows consumers to begin processing immediately once the required data element is available, resulting in a smaller interval between kernels and shorter overall latency.

\textbf{Existing FIFO-based Approaches.} While FIFO is an ideal pattern for large-scale dataflow accelerators, it presents significant challenges for designers. Figure \ref{fig:motivating-exp}(d) shows schedules adopted by Allo \cite{chen2024allo}, StreamHLS \cite{streamhlsFPGA25}, and StreamTensor \cite{streamtensorMICRO25}.
Allo \cite{chen2024allo} presents meaningful attempts to enable FIFO-based dataflow, but two critical issues hinder their on-board execution. First, for the Padding and Conv2D tasks, two  dataflow violations arise (Issue 1). The first violation occurs due to a mismatch of the data access order between adjacent tasks: Padding, as the producer, writes to the FIFO in the loop order of (3,34,34), while Conv2D, the consumer, reads from the FIFO in the loop order of (34,34,3). This discrepancy violates the consistent sequential accessing constraint of FIFOs. The second violation stems from an inconsistency in the data access count between tasks. 
As shown in the right part of Fig. \ref{fig:motivating-exp}, after Padding writes its last data at iteration i, no additional data is written to the FIFO, yet the consumer (Conv2D) is still waiting. This inconsistency leads to a deadlock after iteration i+2. 
Second, existing works, including \cite{chen2024allo}, do not fully explore memory optimization with array partitioning (Issue 2 \ding{172}). For example, in Conv2D shown in the right part of Fig. \ref{fig:motivating-exp}, Allo \cite{chen2024allo} employs a pipeline pragma on the outermost loop, causing all inner loops to be unrolled. However, without proper array partitioning, all $3\times3\times3$ elements are sequentially read from a dual-port BRAM, resulting in a 14-cycle delay. 

StreamTensor \cite{streamtensorMICRO25} introduces a type system for dataflow optimization. When types of a producer and consumer match, FIFO optimization can be performed. However, in cases of mismatch as illustrated in Issue 1, it falls back to suboptimal ping-pong executions rather than resolving underlying violations.
StreamHLS \cite{streamhlsFPGA25} provides partial solutions for Issue 1 by reordering loops and inserting control logic to enable FIFO-based dataflow. 
However, this method does not ensure that FIFO writes start as early as possible, and some cases may require further loop rewriting to achieve effective dataflow.
Consequently, its limited code pattern analysis and transformation lead to suboptimal designs.
First, it fails to identify and enable FIFO-based dataflow between Padding and Conv2D, reverting to ping-pong or sequential execution. 
Second, the data transfer process is inefficient due to a misalignment between the control logic and the loop execution order. As illustrated in Fig. \ref{fig:motivating-exp} (Issue 2, \ding{173}), the control condition \textit{(r==2, c==2, ...)} defers all FIFO write operations until a late stage within the iteration sequence. More precisely, FIFO writes are postponed until nearly 8/9 of the total iterations have completed. This delay introduces substantial pipeline bubbles, leaving the consumer task idle for most of the time.
Quantitative comparison is shown in Fig. \ref{fig:motivating-exp}(b). 


\textbf{CODO Approach.} CODO resolves all the aforementioned issues. It eliminates inconsistencies in data access between adjacent kernels and tasks through advanced code analysis and transformation, and employs communication buffer optimization to 
enable stable and efficient streaming processing on actual FPGA boards. 
In addition, CODO performs automated, resource-aware dataflow scheduling, achieving high parallelism without overusing limited hardware resources, while carefully coordinating neighboring producers and consumers to prevent new dataflow violations. 
In this way, CODO automatically generates highly efficient and deployable dataflow 
accelerators, demonstrating superior effectiveness and practical usability compared to prior methods.

\begin{figure}
    \centering
    \includegraphics[width=0.47\textwidth]{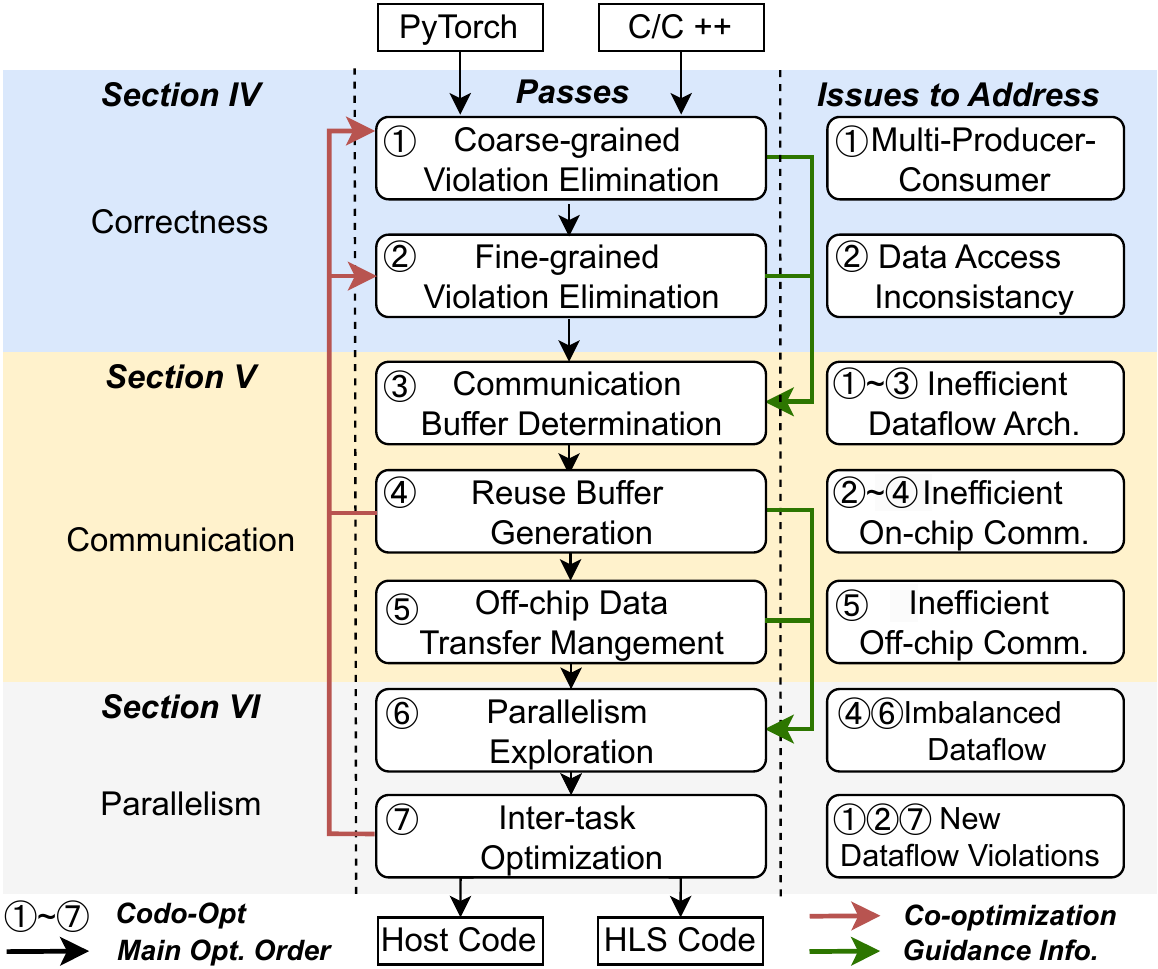}
    \caption{Framework overview of CODO.}
    \vspace{-0.2cm}
    \label{fig:framework}
    \vspace{-0.4cm}
\end{figure}

\section{Framework Overview} \label{Framework Overview}
CODO is built on the MLIR \cite{lattner2020mlir} compilation framework. Figure \ref{fig:framework} shows the compilation flow. The framework takes compute kernels implemented in C++ or PyTorch models as input, which are translated into MLIR dialects via Polygeist \cite{polygeistPACT} and Torch-MLIR \cite{torchmlir}, respectively. CODO offers \texttt{codo-opt}, which applies the full optimization flow in a single command, allowing users to optionally adjust input parameters like maximum parallelism and tiling factors.

CODO contains a holistic compilation flow that follows a main optimization order while being deeply integrated through co-optimization.
The flow begins with two dataflow correction passes. The coarse-grained violation elimination resolves single-producer-consumer violations between tasks, where each task is represented as a \textit{node} in the dataflow graph. Subsequently, the fine-grained violation elimination fixes inconsistencies in data access order and count, enabling efficient FIFO-based communication.
This pass exemplifies our co-optimization principle: beyond ensuring correctness, it proactively restructures code for communication efficiency and provides guidance for later communication passes.
Based on this, CODO performs communication buffer determination, selecting either FIFO or ping-pong implementations and prioritizing FIFO whenever feasible for higher performance.
To further improve communication efficiency, CODO generates efficient reuse buffers and reinvokes the correctness passes to avoid new violations. This process also exposes loop-level parallelism, providing key information for subsequent parallelism exploration.
Afterward, CODO manages off-chip transfers to improve HBM bandwidth utilization. 
Finally, CODO's auto-scheduling engine determines tiling factors, unroll factors, pipelining, and array partitioning. These parallelism decisions are not made in isolation, as they can affect both correctness and communication efficiency. Therefore, a final inter-task optimization pass co-optimizes these choices across the entire graph, eliminating any newly introduced violations and ensuring a high-performance design. 

\section{Dataflow Violation Elimination}
Commercial HLS tools \cite{vitishls,intelhls} exhibit limitations in effectively addressing dataflow violations. These tools only report coarse-grained dataflow violations through synthesis analysis and cannot automatically transform the code to resolve violations. To address these issues, we systematically eliminate both coarse-grained and fine-grained dataflow violations.
 \vspace{-0.1cm}
\subsection{Coarse-grained Violation Elimination} \label{sec:coarse}

\begin{figure}
    \centering
    \includegraphics[width=0.45\textwidth]{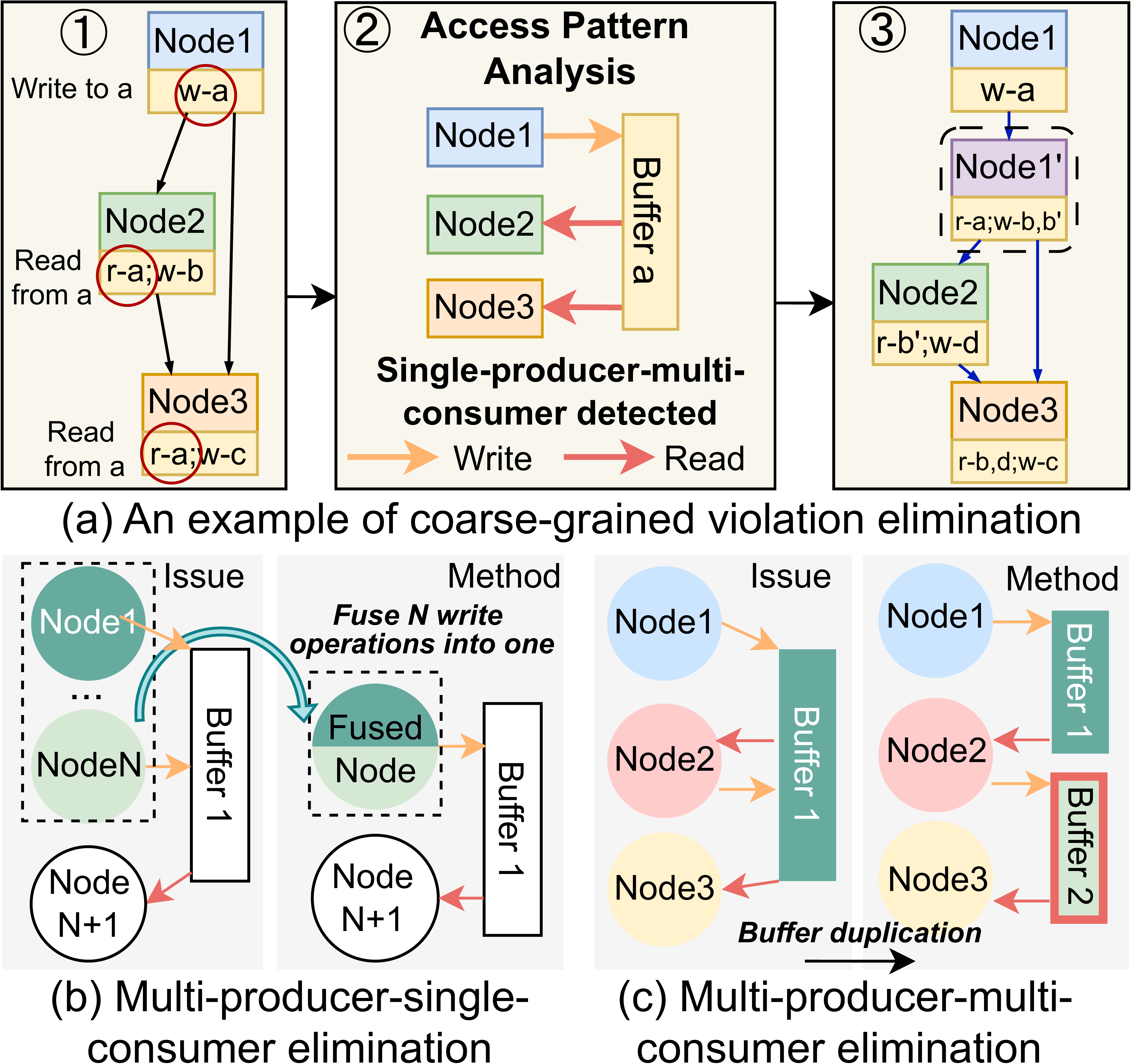}
    \vspace{-0.1cm}
    \caption{Coarse-grained dataflow violation elimination: (a) an example, where r-x/w-x represent read from/write to buffer x, and Node1' represents the node inserted to eliminate the violation; (b)(c) illustrate our elimination techniques for the rest of the dataflow violation categories.}
    \label{fig:corse_eli}
    \vspace{-0.3cm}
\end{figure}

\textbf{Violation Issues.} The input C/C++ code or PyTorch models are first translated into a dataflow graph, where nodes represent computational tasks such as loops or functions, as shown in Fig. \ref{fig:corse_eli}. 
Existing commercial HLS tools enforce a single-producer-single-consumer pattern for dataflow execution, as discussed in Section \ref{sec:background-violation}. Therefore, effective techniques are necessary to eliminate violations that deviate from this constraint.
Figure \ref{fig:corse_eli} illustrates different types of coarse-grained dataflow violations. For example, in Fig. \ref{fig:corse_eli}(a), \textit{Node1} writes results to buffer \texttt{a}, while both \textit{Node2} and \textit{Node3} read from the same buffer, forming a single-producer-multi-consumer pattern. 
Similarly, Fig. \ref{fig:corse_eli}(b) and Fig. \ref{fig:corse_eli}(c) depict multi-producer-single-consumer and multi-producer-multi-consumer patterns, respectively. Although previous works \cite{ye2022scalehls, ye2024hida, streamhlsFPGA25} partially address violations (a) or (c), they often fail to eliminate all violations, leading to sequential execution between nodes.

\textbf{Pattern-aware Code Transformation.} 
To fully address these issues, we propose pattern-aware code transformation, as described in Algorithm \ref{alg:template_compact}. The algorithm traverses the input code and detects data access patterns that may lead to violations (L3-4), which arise when \textit{multiple nodes access the same buffer}. In general, all the access patterns that cause coarse-grained dataflow violations can be classified into three categories, as shown in Fig. \ref{fig:corse_eli}. Once a violation is identified, CODO detects its access pattern
and applies corresponding transformations to refactor the code (L5-6). 
For instance, Fig. \ref{fig:corse_eli}(a)\ding{172} illustrates a typical bypass pattern, commonly seen in models with residual structures such as ResNet-18 \cite{he2016deep} and GPT-2 \cite{Radford2019LanguageMA}. CODO begins by traversing all buffers and collecting all nodes that access them. Taking buffer \texttt{a} as an example, its relevant nodes are \textit{Node1-3}. CODO analyzes and records the access behavior of each node for buffer \texttt{a} (Fig. \ref{fig:corse_eli}(a)\ding{173}), which is then identified as the single-producer-multiple-consumer pattern.
To resolve this violation, 
an intermediate node (\textit{Node1'}) is inserted, reading
from buffer \texttt{a} and writing to duplicated buffers \texttt{b} and \texttt{b'} (Fig. \ref{fig:corse_eli}(a)\ding{174}).

CODO applies different code transformations to address all three coarse-grained violation patterns in Fig. \ref{fig:corse_eli}. The \textit{multi-producer-single-consumer} pattern in Fig. \ref{fig:corse_eli}(b), often found in initialization and padding operation pairs, is resolved through \textit{node fusion}. 
CODO fuses loops that write to the same buffer when they share the same outer iteration domain and have no loop-carried dependencies. If inner loop structures differ, additional control logic is inserted to handle the mismatch.
To maintain correctness, intermediate results from earlier writes are temporarily stored and finally merged into the last write operation.
For the multi-producer-multi-consumer issue in Fig. \ref{fig:corse_eli}(c), we create a new \textit{buffer2} by duplicating \textit{buffer1}, ensuring that each buffer is read from and written to once. 

{\tiny
\renewcommand{\baselinestretch}{1}
\begin{algorithm}[t]
    \renewcommand{\algorithmicrequire}{\textbf{Input:}}
    \renewcommand{\algorithmicensure}{\textbf{Output:}}
    \caption{Pattern-aware Violation Elimination}
    \label{alg:template_compact}
    \begin{algorithmic}[1]
        \REQUIRE Initial input code $\mathbf{M}$ with nodes and buffers.
        \ENSURE Transformed code $\mathbf{M^{'}}$ without violations.

        \STATE \textbf{for all} $buf \in \mathbf{M}$:
        \STATE \quad Collect all nodes $\mathbf{N}$ that access $buf$.
        \STATE \quad $\mathbf{V} \gets \text{analyze\_access\_pattern}(\mathbf{N})$
        \STATE \quad \textbf{if} $\mathbf{V}$ contains violations:
        \STATE \quad \quad Detect the data access pattern $\mathbf{P}$.
        \STATE \quad \quad $\mathbf{N^{'}} \gets \text{apply\_transformation}(\mathbf{N}, \mathbf{P})$
        \STATE \quad \quad $\mathbf{M^{'}} \gets \text{update\_affected\_nodes}(\mathbf{M}, \mathbf{N^{'}})$

        \STATE \textbf{return} $\mathbf{M^{'}}$.
    \end{algorithmic}
\end{algorithm}
}
\vspace{-0.2cm}



\subsection{Fine-grained Violation Elimination}\label{Fine-grained Violation Elimination}
\label{sec:fine-grain vio}
After the coarse-grained violation elimination, HLS tools can by default allocate ping-pong buffers between nodes to enable coarse-grained dataflow execution with data blocks.
However, the performance may not be maximized at nodes whose input and output data can be transferred through FIFOs in sequential order and processed at a finer granularity. This is because FIFO-based dataflow often offers superior performance due to its streaming computation pattern and less resource overhead. However, 
it imposes strict requirements on code patterns, requiring fine-grained violation elimination.

\textbf{Violation Issues.}
In the example of Fig. \ref{fig:corse_eli}(a)\ding{174}, FIFOs can be inserted at all the connections between nodes or loops within nodes, only if the sequential data access constraint is satisfied and the data access orders/counts are consistent between adjacent nodes or loops. 
Unfortunately, real-world applications exhibit numerous fine-grained read-write inconsistencies. As discussed in Section \ref{sec:moti}, violations such as \textit{access count mismatch} and \textit{access order inconsistency} can result in deadlock or computational errors in the final design. 
More critically, existing HLS tools cannot detect these violations during synthesis. While a subset of issues may be identified through cosimulation, the process is time-consuming, often taking days or even weeks, thereby significantly increasing the debugging burden for developers. 

\begin{figure}
    \centering
    \includegraphics[width=0.48\textwidth]{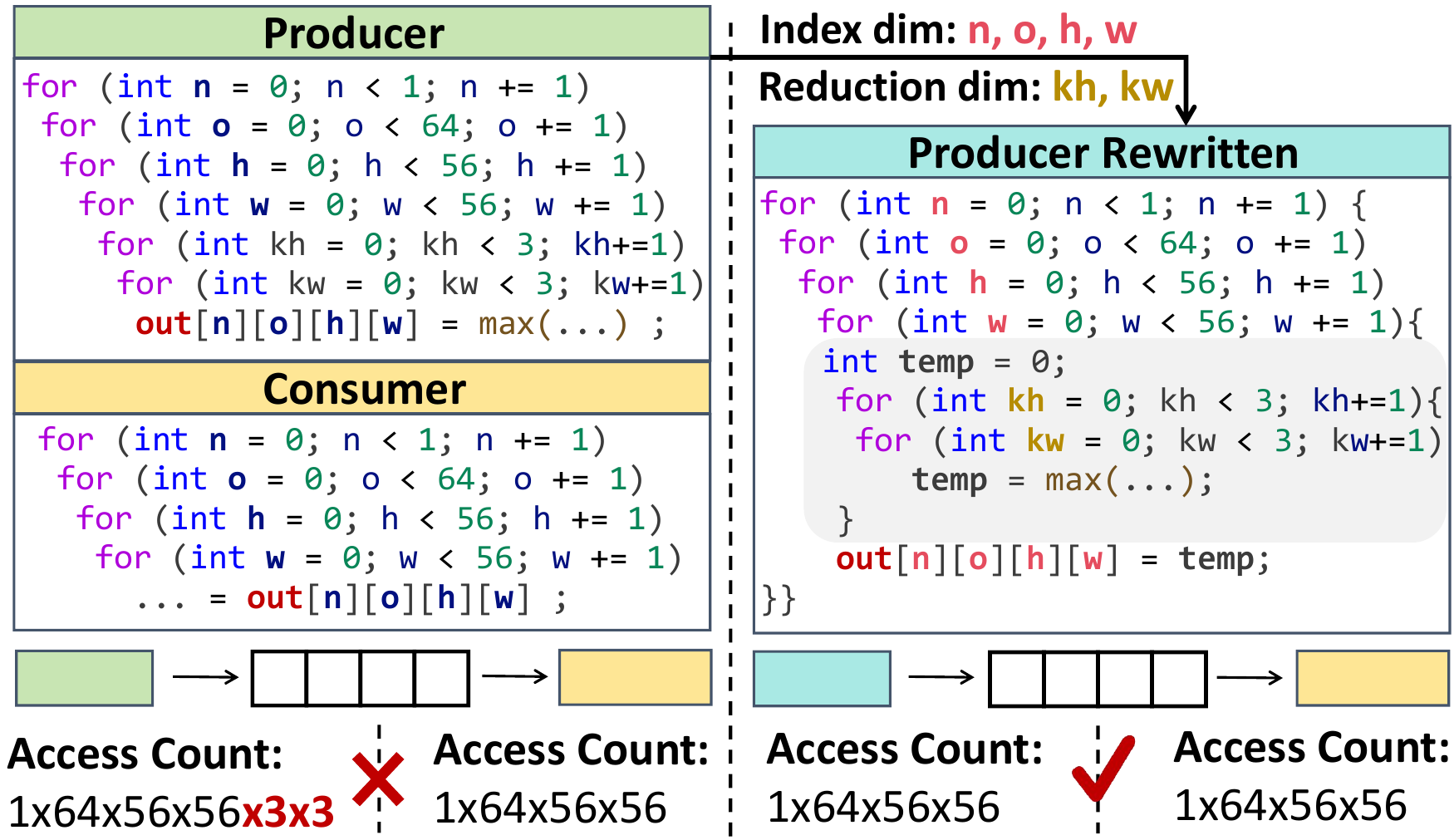}
    \vspace{-0.2cm}
    \caption{An example of a reduction operation rewriting.}
    \vspace{-0.4cm}
    \label{fig:reduction}
\end{figure}

\textbf{Systematic Read-Write Coordination.}
To address fine-grained violations and further refine the design for higher efficiency, we introduce a systematic read-write coordination
method, including 1) \textit{reduction operation rewriting}, which resolves the \textit{data access count mismatch issue while guaranteeing early FIFO writes}, and  
2) \textit{permutation map generation}, which adjusts the access pattern of adjacent loops and ensures their consistent data access order, resolving the \textit{data access order inconsistency} issue.

\textit{1) Reduction Operation Rewriting.} 
Most data access count mismatches stem from reduction operations, such as fully connected layers, max pooling, and normalization. These non-bottleneck operations introduce loop dimensions that do not directly correspond to array indices, resulting in redundant FIFO accesses during reduction iterations. 
To address this issue, we propose a \textit{reduction rewriting strategy} that identifies reduction regions and utilizes temporary arrays to aggregate intermediate results, thereby minimizing unnecessary FIFO transactions and ensuring correct and efficient dataflow execution.

Figure \ref{fig:reduction} illustrates a FIFO access mismatch and our approach. In this example, the producer is a max pooling operation that writes to buffer \texttt{out}, while the consumer is an initialization operation that reads from the same buffer. 
A discrepancy between the number of writes and reads results in a data access count mismatch, which leads to a FIFO deadlock. 
To detect such cases, CODO analyzes the loop structures of both the producer and the consumer that access the same array. It determines the total number of writes and reads by identifying the loop level at which the target array is accessed and computing the product of the iteration counts of the surrounding loops. When a mismatch is detected, CODO classifies loop dimensions that correspond to FIFO array indices as \textit{index dimensions}, while the remaining ones are identified as \textit{reduction dimensions} and moved to the innermost loops, as shown in the shaded region of Fig. \ref{fig:reduction}. The write to \texttt{out} is then moved out of the reduction region, and a temporary buffer is introduced to accumulate intermediate results. This transformation ensures that the producer's access count matches that of the consumer. Moreover, the rewriting ensures that intermediate results are being calculated and transferred just-in-time, greatly improving data transfer efficiency.

\begin{figure}
    \centering
    \includegraphics[width=0.48\textwidth]{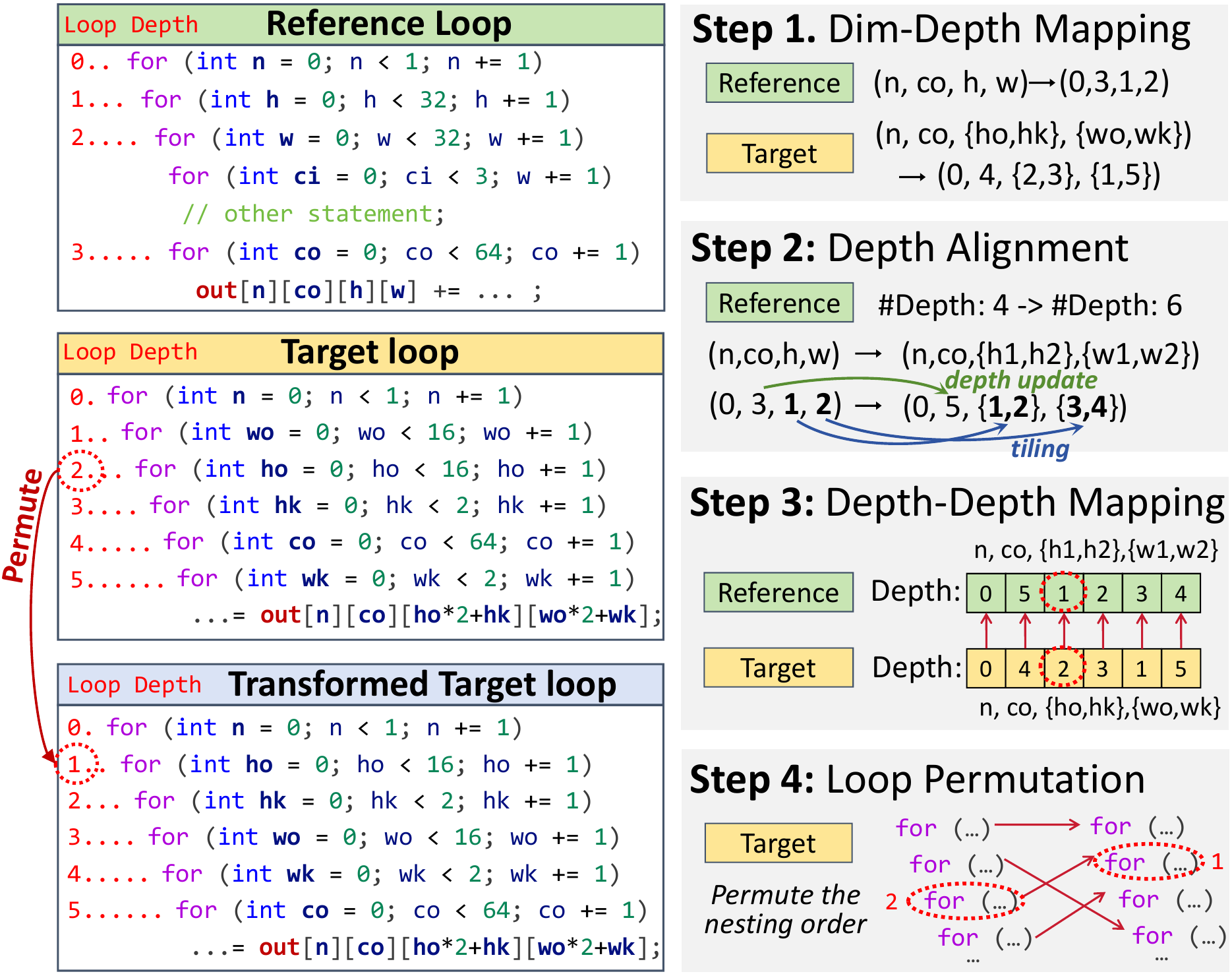}
    \vspace{-0.2cm}
    \caption{Illustration of permutation map generation.}
    \vspace{-0.4cm}
    \label{fig:permutation}
\end{figure}

\textit{2) Permutation Map Generation.} 
Inconsistent data access orders, 
which are common in real-world applications, lead to dataflow violations for streaming processing with FIFOs, as illustrated in Issue 1 of Fig. \ref{fig:motivating-exp}.
To address this issue, we propose \textit{permutation map generation}.
Specifically, CODO identifies the bottleneck loop (e.g., convolution or Q*K in attention) as the \textit{reference loop} by analyzing the trip counts and computational intensity of each nested loop.
It then analyzes data access patterns of the \textit{reference loop}, including the data access order of input and output arrays. This information serves as the basis for adjusting the data access patterns of its producer and consumer loops, termed \textit{target loops}. CODO then employs a mapping-based strategy to efficiently align data access patterns between reference and target loops.

Figure \ref{fig:permutation} illustrates this process. In Step 1, CODO establishes a mapping from connection array dimensions to their corresponding loop depths for both reference and target loops. 
For example, in the reference loop, the dimension set \texttt{\{n, co, h, w\}} of \texttt{out} corresponds to the loop depth set \texttt{\{0, 3, 1, 2\}}. In Step 2, we apply loop tiling with a tiling size of 1 to the reference loop to align the depths of the reference loop and the target loop, splitting \texttt{h} and \texttt{w} into two loops, respectively. In Step 3, we construct a mapping between the loop depth sets of the reference and target loops. For instance, a mapping from 2 to 1 indicates that the loop at depth 2 in the target loop should be swapped to depth 1. Finally, in Step 4, we transform the target loop by permuting the nesting order based on the depth-depth map from Step 3. 

\section{Efficient Data Communication} \label{sec:mem}
After eliminating dataflow violations, the input algorithm is transformed into a dataflow-feasible form. 
Based on it, optimizing both on-chip and off-chip data communication is critical for overall efficiency. Therefore, we propose two on-chip optimizations: 1) \textit{communication buffer determination}, which prioritizes FIFOs for tasks without dataflow violations;
2) \textit{violation-free reuse buffer generation}, which enhances data transfer efficiency while ensuring violation-free designs; and an off-chip optimization: 3) \textit{off-chip data transfer management}, which improves HBM bandwidth utilization.

\subsection{On-chip Communication Buffer Determination} \label{sec:dataflow_select}
We adopt a FIFO-first strategy to optimize on-chip communication buffers. For tasks free of violations, we prioritize FIFO implementations to maximize performance. 
If fine-grained violations between loops cannot be eliminated, we turn to ping-pong buffer implementations. 
Note that ping-pong buffers are more resource-intensive, as they require at least twice the buffer size of the transmitted data block, posing a risk of resource overflow in large-scale applications. 

\subsection{Reuse Buffer Generation}\label{Reuse Buffer Generation}
\label{sec:reuse}
Leveraging reuse buffers is effective to enhance data reuse and memory bandwidth, but integrating them into the design requires careful loop refactoring.
Existing HLS frameworks often rely on user-defined DSLs, requiring developers to manually specify buffer sizes, communication buffer types, and address mappings for reused data \cite{lai2019heterocl, xiang2022heteroflow, chen2024allo}. This approach demands developer expertise
and often leads to suboptimal or infeasible designs,
as discussed in Section \ref{sec:moti}. 
To this end,
we propose a \textit{violation-free reuse buffer generation} method that automatically exploits data reuse opportunities.

\begin{figure}
    \centering
    \includegraphics[width=0.48\textwidth]{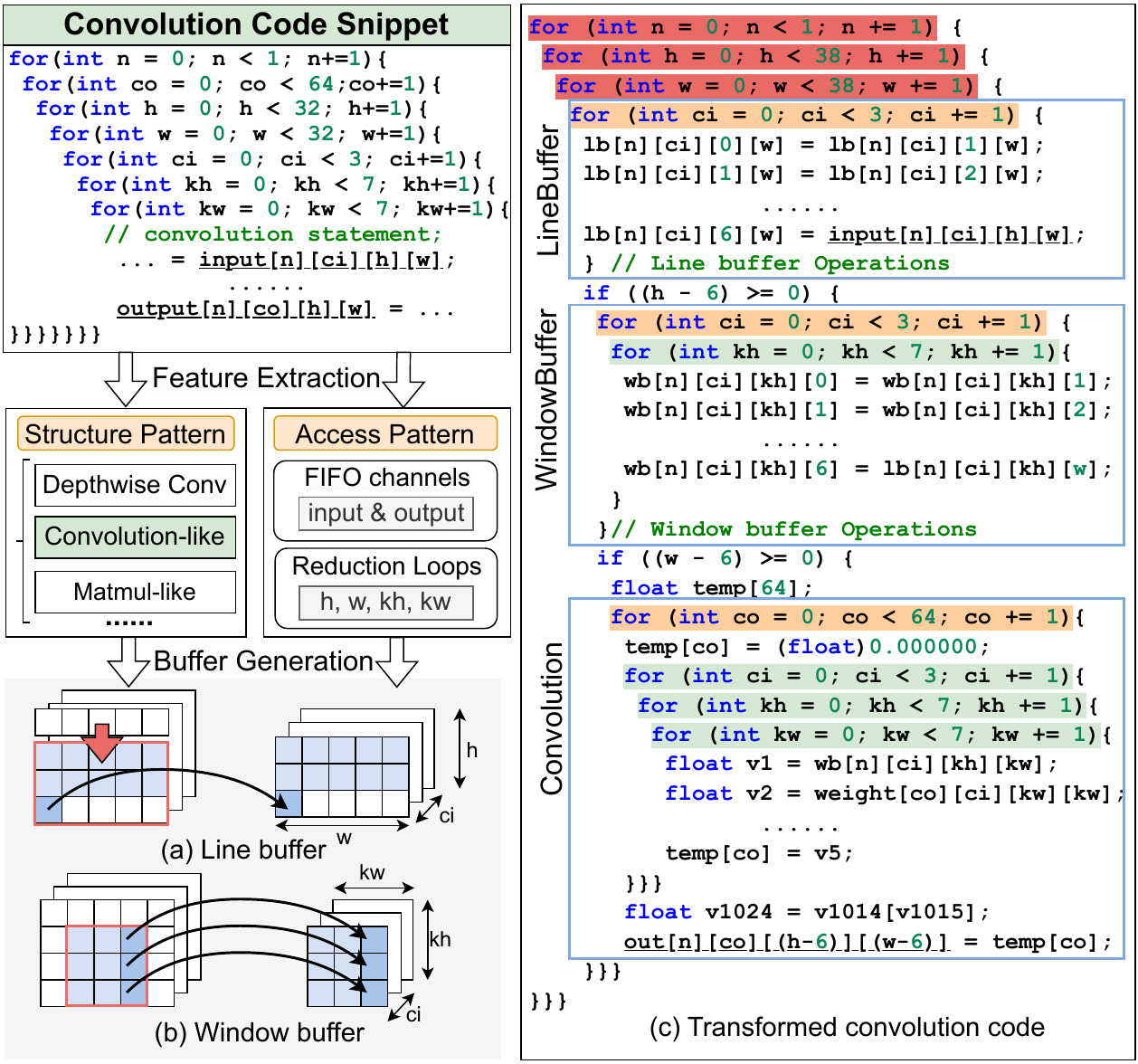}
    \vspace{-0.2cm}
    \caption{Example code for efficient reuse buffer generation.}
    \vspace{-0.4cm}
    \label{fig:reuse-buffer}
\end{figure}

\textbf{Violation-free Reuse Buffer Generation.} 
To automatically generate efficient reuse buffers, 
CODO first analyzes the nested loop structure and the operations within each loop to identify computation-intensive kernels such as convolution and matrix multiplication. This is achieved by detecting common computation patterns, such as multiply-accumulate operations. It then extracts the input/output access patterns of the target array and analyzes the mapping between loop variables and array indices. Loop dimensions that appear in the array indices are identified as FIFO dimensions, while the remaining loop dimensions are treated as reduction dimensions. This information is guidance for subsequent reuse buffer generation.
Taking the convolution example in Fig. \ref{fig:reuse-buffer}, each output pixel depends on a small local region of the input feature map, and neighboring outputs reuse many input elements. These reuse opportunities align with the reduction dimensions. CODO automatically analyzes FIFO indices, identifies reduction dimensions that are independent of the FIFO, and constructs reuse buffers accordingly.
Specifically, CODO constructs line and window buffers based on the iteration domain of reduction loops (Fig. \ref{fig:reuse-buffer}(a),(b)).
The line buffer, denoted as \texttt{lb[n][ci][kh][w]}, stores multiple rows of the input feature map.
Its depth is equal to the kernel height (\texttt{kh}), retaining \texttt{kh-1} rows to preserve history for subsequent computations. Each new input element (\texttt{input[n][ci][h][w]}) is written into the most recent position. The window buffer, denoted as \texttt{wb[n][ci][kh][kw]}, maintains the full \texttt{kh×kw} window of the convolution kernel. For each new column \texttt{w} of the input, it updates by shifting existing contents horizontally and loading the new column from the line buffer. To prevent dataflow violations, CODO refactors loops while analyzing access patterns of FIFOs, ensuring that all loop dimensions are properly utilized. Specifically, loops involving FIFO accesses must neither include irrelevant dimensions nor omit necessary ones. For example, in Fig. \ref{fig:reuse-buffer} (c), the \textit{input} and \textit{output} arrays are optimized as FIFOs, and the nested loops enclosing them precisely align with the array indices, ensuring consistent data accesses. Note that this method is also applicable when the target array is implemented using ping-pong buffers.
\color{black}

\textbf{Guidance for Parallelism Exploration.} 
After reuse buffer generation, the rewritten code is ready for further optimizations through loop tiling, pipelining, unrolling, and array partitioning. However, as shown in Fig. \ref{fig:reuse-buffer}, the generated loop is highly complex, making optimization challenging. By analyzing the internal computation behavior, we identify distinct parallelism opportunities. First, parallelizing the outermost red loop is unsafe, as it would unroll all three internal regions, introducing complex data dependencies and control issues. Second, the middle orange loops are associated with FIFO indices, and optimizing them could alter FIFO access patterns, potentially causing new violations. Finally, the innermost green loops are independent of FIFO behavior, making them safe for parallelization without introducing new violations.
Based on the above analysis, a loop is legal to parallelize if it has no loop-carried dependencies. If the target loop variable appears in FIFO indices, parallelization remains feasible, but additional measures are needed to preserve the consistency of the data access pattern between producer and consumer.
This analysis is crucial for the subsequent parallelism exploration, as it enables the effective pruning of the vast design space associated with large-scale models. 

\subsection{Off-chip Data Transfer Management} \label{sec:off-chip}
To improve off-chip bandwidth utilization, CODO automatically constructs efficient burst transfers between HBM and on-chip memory. It distributes parameters such as model weights across different HBM channels, enabling parallel access to independent memory regions. 
CODO provides a \texttt{codo-transmit} command, which automatically generates the host code and burst-access operations for kernels and users
can specify the number of HBM channels allocated.

\section{Automated Dataflow Scheduling} \label{sec:dse}
After applying previous passes, dataflow violations are eliminated and buffers are inserted at suitable positions. Then CODO performs auto-scheduling to exploit parallelism without exceeding resource budgets and coordinate adjacent tasks without introducing new dataflow violations.

\begin{figure}
    \centering
    \includegraphics[width=1.\columnwidth]{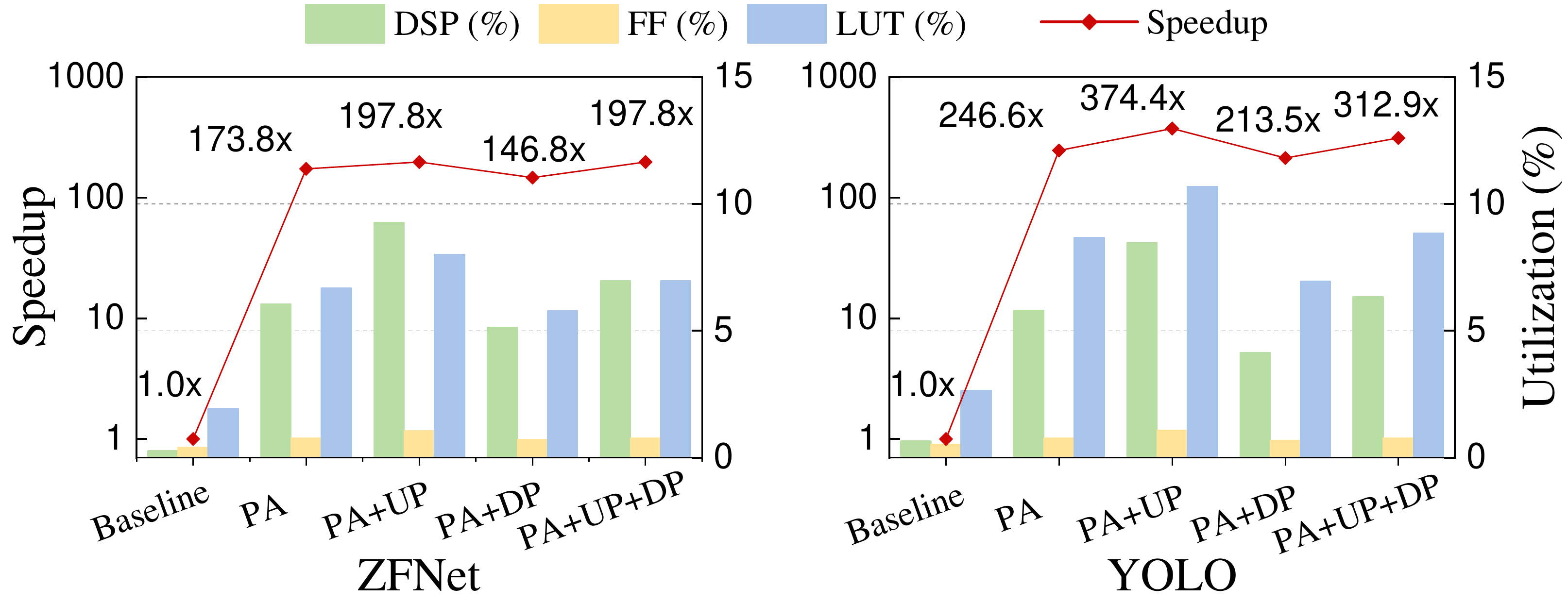}
    \vspace{-0.3cm}
    \caption{Speed and resource util. 
    of parallelism exploration.}
    \vspace{-0.4cm}
    \label{fig:dse-ablation}
\end{figure}

\textbf{Challenges.} Parallelism exploration in dataflow accelerators presents significant challenges: 1) improper parallelism strategies can disrupt latency balance between tasks, leading to degraded dataflow performance; 2) existing methods prioritize performance gains while neglecting the resource-performance tradeoff, which can result in excessive resource consumption; 3) certain parallelization strategies may alter FIFO access patterns, introducing new dataflow violations.


\textbf{Parallelism Exploration.}
To address these challenges, we propose resource-aware bottleneck-centric design space exploration (DSE) to find optimal parallelism strategies, including loop tiling, pipelining, unrolling, and array partitioning configurations.
Fig. \ref{fig:dse-ablation} illustrates the  speedup and resource utilization at each stage of the parallelism exploration process.

\textit{1) Stage One: Initial Parallelism Allocation (\textit{PA}).} 
The DSE begins by constructing a high-quality initial design with initial parallelism degrees. 
CODO employs a profiling-based performance model \cite{zhao2017comba} \cite{ye2022scalehls}. Latencies and resource consumption of basic operations, such as adders, are profiled, serving as the performance model parameters. Then the latency of each loop can be estimated based on their loop trip counts and parallelism strategies. 
After estimating the latency of each loop, CODO allocates parallelism degrees in proportion to their latencies, setting the smallest degree to 1. It then gradually scales up the parallelism of all loops while preserving their proportional ratios until reaching the user-specified upper bound or hardware resource limits. This process helps form a roughly balanced dataflow structure.

Unlike methods that parallelize loop dimensions randomly, CODO leverages insights from the earlier communication optimization pass. It prioritizes tiling loop dimensions that are independent of FIFO accesses, ensuring correct and efficient communication, and automatically applies HLS pragmas such as pipelining, unrolling, and array partitioning to generate the initial design.
As shown in Fig. \ref{fig:dse-ablation}, this stage brings significant benefits,
delivering latency speedups of $173.8 \times$ on ZFNet \cite{zeiler2014visualizing} and $246.6 \times$ on YOLO \cite{YOLO}.



\textit{2) Stage Two: Upscaling (\textit{UP}).} We then traverse all \textit{bottleneck loops} and re-estimate their latencies. If one loop's latency remains at least \( n \) times larger than the lowest loop latency, we increase its parallelism degree to
$\max\{\lceil n \rceil \times \text{initial degree}, \text{max degree}\}$ and search corresponding parallelism strategies.
This process is iterative and terminates when the parallelism degree of all loops stabilizes or the iteration limit is reached. Note that this step aims to maximize the parallelism degree for each bottleneck loop, which may increase resource usage. As shown in Fig. \ref{fig:dse-ablation}, after \textit{PA+UP}, the performance of both models is improved at the cost of extra resource consumption.

\textit{3) Stage Three: Downscaling (\textit{DP}).} Since the overall performance of a FIFO-based dataflow depends on the loop with the longest latency, we introduce a downscaling step to adjust over-optimized loops. If a loop is \( n \) times faster than the longest loop,
it indicates that this loop has been over-optimized. 
In this case, we decrease the parallelism degree of the loop by a factor of \( n \), reducing resource usage while maintaining a balanced dataflow. This step may slightly reduce overall performance, but users can configure CODO to enable or disable it based on their design goals.  As shown in Fig. \ref{fig:dse-ablation}, after applying \textit{PA+UP+DP}, ZFNet achieves the same speedup as the \textit{PA+UP} design but with significantly lower resource consumption. Users can also bypass the \textit{UP} stage and directly perform \textit{DP} to pursue the most resource-efficient design, particularly in resource-constrained scenarios.

The parameter \( n \) acts as a balancing threshold in the parallelism exploration process. In practice, increasing loop parallelism through transformations such as loop unrolling typically has a minimum granularity of 2. Therefore, we empirically set \( n = 2.0 \) to avoid skipping potentially optimal design points that might be missed if larger values (e.g., 3, 4, or 8) were used. We also observe that small variations around this value do not lead to noticeable performance differences. Developers can specify a customized value for  \( n \) in \texttt{codo-opt} to adjust the exploration granularity based on their own design constraints.

\textbf{Inter-task Optimization.}
After parallelism exploration, the bottleneck loops are effectively optimized. However, if loop tiling is applied to dimensions used by FIFO, the loops connected to the other end of FIFO need to adopt the same parallelism strategy to maintain consistent data access behavior. 
CODO implements this by propagating the chosen loop tiling, unrolling, and array partitioning strategies of the bottleneck loop to its connected producer/consumer loops.
These changes often reshape the loop structure, so we re-invoke our correctness passes to detect and resolve any newly introduced dataflow violations.

However, conflicts can arise during this stage. For example, consider a dataflow consisting of loops A, B, C, and D connected by FIFOs. If loops B and D adopt conflicting parallelism strategies, loop C may encounter an unresolvable violation. In such cases, CODO downgrades the buffer between loops C and D to a ping-pong buffer implementation, preserving the FIFO-based execution from loop A to loop C. 



\section{Implementation}
\subsection{Frontend}
CODO is seamlessly integrated into the MLIR infrastructure and supports two primary input pathways: (1) Polygeist \cite{polygeistPACT} for translating C/C++ kernels directly into the affine dialect, and (2) Torch-MLIR \cite{torchmlir} for importing PyTorch models. In the Torch-MLIR flow, models are first lowered to the linalg dialect, where built-in MLIR passes such as element-wise operator fusion and tensor bufferization are applied. The IR is then further lowered to the affine dialect, which serves as the primary representation for subsequent CODO optimizations. CODO targets affine programs with constant loop bounds, which covers a wide variety of kernels and layers in DNNs, linear algebra, and image processing, such as convolution, attention, activation layers (ReLU, GeLU), matrix multiplication, dot product, etc.

\subsection{Optimization Passes}
All optimizations are implemented as modular MLIR passes, enabling high extensibility, as illustrated in Fig. \ref{fig:framework}. After each transformation, MLIR’s built-in verification checks IR validity, including dominance relations, SSA consistency, and type correctness.   In addition, the \texttt{canonicalize} and \texttt{cse} (common sub-expression elimination) passes remove dead code and redundant computations \cite{mlir2025}. Therefore, the MLIR infrastructure inherently guarantees IR validity throughout the compilation process.
Loop transformation passes, including violation elimination and reuse buffer generation, operate primarily on the affine dialect. To optimize data communication, CODO introduces dedicated data types and operations to explicitly model FIFO and ping-pong buffers. Hardware-specific semantics, such as dataflow pragmas and array partitioning directives, are represented through an extended HLS dialect based on the version originally developed by HIDA \cite{ye2024hida}.

\subsection{Translation}
After optimizations, the transformed IR is lowered to the host code and the HLS C++ kernel code. To ensure functional correctness, we adopt the verification pipeline from StreamHLS \cite{streamhlsFPGA25}, which automatically generates a testbench to validate functional equivalence by comparing the accelerator outputs with the golden results of the original program.

\section{Experiments}
\textbf{Setup.} We evaluate performance with Xilinx Vitis HLS and Vivado 2023.2 for synthesis and hardware implementation. Latency and resource statistics are collected from HLS synthesis reports, and runtime and power statistics are measured through on-board evaluation.
The platform is an AMD Alveo U280 FPGA board, containing 9024 DSP slices, 2.6M flip-flops, 1.3M LUTs, and 4032 BRAM18K blocks. The target frequency is set to 300 MHz for all experiments.

\textbf{Baseline and Workload.} We compare CODO with six compilers, including ScaleHLS \cite{ye2022scalehls}, POM \cite{zhang2024pom}, Allo \cite{chen2024allo}, HIDA \cite{ye2024hida}, StreamHLS \cite{streamhlsFPGA25}, and StreamTensor \cite{streamtensorMICRO25}, and an accelerator, DFX \cite{dfxmicro2022}.
We begin by evaluating performance on typical kernel-level applications from PolyBench and widely used models \cite{encoder2010,he2016deep,howard2017MobileNets,vaswani2017attention}.
We further compare these frameworks on more complex DNN workloads, including ResNet-18 \cite{he2016deep}, VGG-16 \cite{simonyan2014very}, MobileNet \cite{howard2017MobileNets}, ZFNet \cite{zeiler2014visualizing}, and YOLO\cite{YOLO}. To demonstrate CODO's applicability to large language models (LLMs), we also compare CODO against other frameworks on a transformer-based model, GPT-2 \cite{Radford2019LanguageMA}.

\begin{table}[t]
    \renewcommand{\arraystretch}{1.0}
    \centering
    \caption{Evaluation on typical kernel-level applications.}
    \label{tab:polybench-syn}
    \resizebox{\columnwidth}{!}{%
    \begin{tabular}{@{}cccccc@{}}
    \toprule
    \multirow{2}{*}{\textbf{Benchmark}} & 
    \multirow{2}{*}{\textbf{DSP}} & 
    \multicolumn{4}{c}{\textbf{Latency Speedup}} \\
    \cmidrule(lr){3-6}
    & & \textbf{CODO} & \textbf{StreamHLS} & \textbf{Allo} & \textbf{HIDA} \\
    \midrule
    Atax & 602 & \textbf{853.3$\times$} & 640.1$\times$ & 293.2$\times$ & 11.1$\times$ \\
    Gesummv & 562 & 369.1$\times$ & \textbf{382.4$\times$} & 329.1$\times$ & 79.8$\times$ \\
    Gemm & 826 & 500.8$\times$ & \textbf{600.4$\times$} & 177.4$\times$ & 239.2$\times$ \\
    Mvt & 600 & \textbf{488.1$\times$} & 368.7$\times$ & 249.5$\times$ & 79.7$\times$ \\
    3mm & 830 & 379.0$\times$ & \textbf{442.3$\times$} & 321.0$\times$ & 180.5$\times$ \\ 
    \hline
    Residual MLP & 648 & \textbf{449.2$\times$} & 449.2$\times$ & -- & -- \\
    Autoencoder & 696 & \textbf{329.4$\times$} & 329.3$\times$ & 7.7$\times$ & 141.4$\times$ \\
    Residual Block & 488 & \textbf{225.8$\times$} & 99.1$\times$ & -- & 127.7$\times$ \\
    DWSConv. Block & 495 & \textbf{14.4$\times$} & 7.7$\times$ & -- & -- \\
    3-Layer Conv. Block & 613 & \textbf{209.0$\times$} & 37.2$\times$ & 122.1$\times$ & -- \\ 
    Feed Forward & 604 & \textbf{513.2$\times$} & 256.6$\times$ & 3.8$\times$ & -- \\
    Multi-Head Attention & 848 & \textbf{256.5$\times$} & 168.5$\times$ & 4.0$\times$ & -- \\
    \midrule
    \textbf{DSE time} & & (0.1s-0.5s) &(35s-20min) & - & (0.4s-5min32s) \\
    
    \midrule
    \textbf{Geo. Mean} & & \textbf{292.1$\times$} & 200.9$\times$ & 64.7$\times$ & 91.8$\times$ \\
    \bottomrule
    \end{tabular}%
    }
    \vspace{-0.5cm}
\end{table}

{\fontsize{15}{11}\selectfont
\begin{table*}[ht]
    \renewcommand{\arraystretch}{1.0}
\caption{Evaluation on different DNN models with input size 3*32*32. 
}
\label{tab:syn small DNN}
\centering
\resizebox{0.95\textwidth}{!}{
\begin{tabular}{ccccccccccc}
\toprule[1pt]
\multirow{2}{*}{\begin{tabular}[c]{@{}c@{}}\textbf{Application} \end{tabular}} &  \multirow{2}{*}{\begin{tabular}[c]{@{}c@{}}\textbf{Framework}\end{tabular}} & \multirow{2}{*}{\begin{tabular}[c]{@{}c@{}} \textbf{Latency} \\ \textbf{(cycles)} \end{tabular}}   &   \multirow{2}{*}{\begin{tabular}[c]{@{}c@{}} \textbf{Speedup} \end{tabular}} &  \multirow{2}{*}{\begin{tabular}[c]{@{}c@{}} \textbf{Compilation}\\ \textbf{Time(s)} \end{tabular}}  & \multirow{2}{*}{\begin{tabular}[c]{@{}c@{}} \textbf{BRAM} \\ \textbf{(Util.\%)} \end{tabular}}  & \multirow{2}{*}{\begin{tabular}[c]{@{}c@{}} \textbf{DSP} \\ \textbf{(Util.\%)} \end{tabular}} & \multirow{2}{*}{\begin{tabular}[c]{@{}c@{}} \textbf{FF} \\ \textbf{(Util.\%)} \end{tabular}} & \multirow{2}{*}{\begin{tabular}[c]{@{}c@{}} \textbf{LUT} \\ \textbf{(Util.\%)} \end{tabular}} \\ \\ 
\toprule[1pt]
\multirow{4}{*}{\begin{tabular}[c]{@{}c@{}}\textbf{ResNet-18}\end{tabular}} 
& ScaleHLS           &  104.88M &   5.3$\times$  & 60.8  & 8416 (208.7\%) &  1330 (14.7\%)    &   144K (5.5\%)   &   992K (76.1\%)   \\ 
& POM & 20.33M &  27.4$\times$  & 77.3 &  0 (0.0\%) & 577 (6.4\%) & 22K (0.9\%)& 90K (6.9\%) \\ 
& Allo &  8.29M  &   66.9$\times$  &  -  &   0  (0.0\%)  &   652 (7.4\%)   &   51K (3.5\%)    &  124K (9.8\%)    \\ 
 &\textbf{CODO} & \textbf{1.69M} & \textbf{326.6$\times$}  & \textbf{1.2} & 116 (2.9\%) & 468 (5.2\%) & 22K (0.9\%) & 103K (7.9\%) \\\hline

\multirow{4}{*}{\begin{tabular}[c]{@{}c@{}}\textbf{VGG-16}\end{tabular}}  
& ScaleHLS           &   28.31M &   6.8$\times$  &  37.3 &  3936 (97.6\%)    &  882 (9.8\%)     & 100K (3.8\%)     &    714K (54.8\%)  \\
& POM  & 10.16M & 18.9$\times$  & 57.8 &  0 (0.0\%) & 416 (4.6\%) &  19K (0.7\%) & 75K (5.7\%)\\ 
& Allo &  3.85M & 50.1$\times$  & - & 0 (0.0\%) & 440 (4.9\%) &  36K (1.4\%)     &    98K (7.5\%)  \\ 
& \textbf{CODO} & \textbf{1.22M} & \textbf{158.0$\times$}  & \textbf{1.0} & 60 (1.7\%) & 376 (4.2\%) & 16K (0.6\%) & 79K (6.1\%) \\ \hline

\multirow{4}{*}{\begin{tabular}[c]{@{}c@{}}\textbf{MobileNet}\end{tabular}}  
& ScaleHLS &   2.17M  &  5.6$\times$  &  38.1  & 6796 (168.6\%) &   1778 (19.7\%)   &    93K (3.6\%)   &   518K (39.7\%)   \\ 
& POM  & 2.02M & 6.0$\times$   & 139.0 &  0 (0.0\%) & 928 (11.7\%) & 33K (1.7\%) & 143K (13.1\%) \\ 
& Allo &  0.26M  & 46.6$\times$  & - & 0 (0.0\%) & 1942(21.5\%) & 57K (2.2\%)    &    128K (9.8\%)      \\ 
& \textbf{CODO} & \textbf{0.08M} & \textbf{151.5$\times$}  & \textbf{1.1} & 70 (1.7\%) & 220 (2.4\%) & 14K (0.6\%) & 62K (4.8\%) \\ 

\bottomrule[1pt]
\end{tabular}
}
\end{table*}
}

{\fontsize{15}{11}\selectfont
\begin{table*}[ht]
    \renewcommand{\arraystretch}{1.0}
\caption{Evaluation on different DNN models with input size 3*224*224 (except YOLO: 3*1280*384). 
}
\label{tab:syn large DNN}
\centering
\resizebox{0.95\textwidth}{!}{
\begin{tabular}{ccccccccccc}
\toprule[1pt]
\multirow{2}{*}{\begin{tabular}[c]{@{}c@{}}\textbf{Application} \end{tabular}} &  \multirow{2}{*}{\begin{tabular}[c]{@{}c@{}}\textbf{Framework}\end{tabular}} & \multirow{2}{*}{\begin{tabular}[c]{@{}c@{}} \textbf{Latency} \\ \textbf{(cycles)} \end{tabular}}   &   \multirow{2}{*}{\begin{tabular}[c]{@{}c@{}} \textbf{Speedup} \end{tabular}} &  \multirow{2}{*}{\begin{tabular}[c]{@{}c@{}} \textbf{Compilation}\\ \textbf{Time(s)} \end{tabular}}  & \multirow{2}{*}{\begin{tabular}[c]{@{}c@{}} \textbf{BRAM} \\ \textbf{(Util.\%)} \end{tabular}}  & \multirow{2}{*}{\begin{tabular}[c]{@{}c@{}} \textbf{DSP} \\ \textbf{(Util.\%)} \end{tabular}} & \multirow{2}{*}{\begin{tabular}[c]{@{}c@{}} \textbf{FF} \\ \textbf{(Util.\%)} \end{tabular}} & \multirow{2}{*}{\begin{tabular}[c]{@{}c@{}} \textbf{LUT} \\ \textbf{(Util.\%)} \end{tabular}} \\ \\ 
\toprule[1pt]
\multirow{2}{*}{\begin{tabular}[c]{@{}c@{}}\textbf{ResNet-18}\end{tabular}}                 
& HIDA & 74.85M & 29.7$\times$  & 83.1 & 1857 (46.1\%) & 574 (6.4\%) & 48K (1.8\%) & 132K (10.2\%) \\ 
& \textbf{CODO} & \textbf{4.76M} & \textbf{466.5$\times$}  & \textbf{1.4} & 548 (13.6\%) & 535 (5.9\%) & 26K (1.0\%) & 115K (8.8\%)   \\ \hline

\multirow{2}{*}{\begin{tabular}[c]{@{}c@{}}\textbf{VGG-16}\end{tabular}}   
& HIDA & 56.93M & 83.0$\times$  & 199.9 & 2344 (58.1\%) & 1163 (12.9\%) & 67K (2.6\%) & 209K (16.0\%)  \\ 
& \textbf{CODO} & \textbf{7.85M} & \textbf{601.8$\times$}  & \textbf{4.3} & 141 (3.5\%) & 952 (10.5\%) & 32K (1.2\%) & 167K (12.8\%)  \\ \hline

\multirow{2}{*}{\begin{tabular}[c]{@{}c@{}}\textbf{MobileNet}\end{tabular}} 
& HIDA & 23.14M & 23.4$\times$  & 110.8 & 2060 (51.1\%) & 782 (8.7\%) & 49K (1.9\%) & 141K (10.8\%) \\ 
& \textbf{CODO} & \textbf{2.02M} & \textbf{268.5$\times$}  & \textbf{1.8} & 256 (6.3\%) & 677 (7.5\%) & 23K (0.9\%) & 123K (9.4\%)  \\ \hline

\multirow{2}{*}{\begin{tabular}[c]{@{}c@{}}\textbf{ZFNet}\end{tabular}}    
& HIDA & 15.49M & 69.9$\times$  & 116.2 & 1223 (30.3\%) & 644 (7.1\%) & 31K (1.2\%)  & 83K (6.4\%)    \\ 
& \textbf{CODO} & \textbf{5.48M} & \textbf{197.7$\times$}  & \textbf{6.1} & 68 (1.7\%) & 630 (7.0\%) & 20K (0.8\%) & 91K (7.0\%) \\ \hline

\multirow{2}{*}{\begin{tabular}[c]{@{}c@{}}\textbf{YOLO}\end{tabular}}    
& HIDA & 59.64M & 83.9$\times$  & 188.2 & 1989 (49.3\%) & 919 (10.2\%)  & 50K (1.9\%) & 152K (11.7\%)   \\ 
& \textbf{CODO} & \textbf{11.90M} & \textbf{420.5$\times$}  & \textbf{4.4} & 147 (3.6\%) & 1132 (12.5\%) & 53K (2.0\%) & 171K (13.1\%) \\ 
\bottomrule[1pt]
\end{tabular}
}
\vspace{-0.3cm}
\end{table*}
}




\subsection{Evaluation on Typical Kernel-level Applications}
The comparison is shown in Table \ref{tab:polybench-syn}, presenting the \textit{latency speedup} 
along with the resource usage of CODO. The \textit{latency speedup} is computed through dividing the latency (\textit{\#clock cycles}) of the Vitis HLS-optimized code by that of the framework-optimized code. We set the same resource budget (DSP = 900, about 1/3 of the DSPs of a single super logic region
) for all the frameworks.
Allo, HIDA, and StreamHLS are compared 
because they focus on dataflow optimization, demonstrating higher performance on applications with multiple kernels compared to ScaleHLS and POM. 

For simple kernels with few dataflow optimization opportunities in Polybench, CODO achieves competitive or higher latency speedups.
For more complex deep learning workloads, HIDA delivers unsatisfying performance due to lacking support of FIFO-based dataflow, while Allo suffers from performance degradation due to the lack of automated scheduling. StreamHLS fails to eliminate all violations and generates accelerators with discontinuous dataflow regions, leading to sequential or ping-pong-based execution in applications such as the 3-Layer Block and DWSConv. In contrast, 
CODO extensively eliminates fine-grained violations and performs communication buffer optimization, achieving $1.45 \times$, $4.52\times$, and $3.18 \times$ latency speedups on average compared to StreamHLS, Allo, and HIDA, respectively. For DSE time, StreamHLS's MINLP solver takes 
exponentially increased search time as the application complexity grows,
whereas CODO only takes seconds to find a high-performance design for all applications.

\subsection{Evaluation on DNN Models}\label{Evaluation on CNN Models}
\label{sec:dnn}
We compare CODO with SOTA frameworks on various DNN models.
For fair comparison, we use the same input sizes as in ScaleHLS, POM, and Allo (3*32*32) and extend the comparison with HIDA to ZFNet and YOLO (3*224*224). StreamHLS fails to generate valid designs for large models, and StreamTensor is excluded as it only targets LLM acceleration.
The results are shown in Table \ref{tab:syn small DNN} and \ref{tab:syn large DNN}. 

\textbf{Performance Comparison.} 
CODO significantly outperforms existing frameworks, achieving average speedups of $33.8\times$, $13.6\times$, $3.7\times$, and $7.1\times$ over ScaleHLS, POM, Allo, and HIDA, respectively, while also reducing compilation time, defined as the time required for all optimizations, DSE, and code generation. The improvement stems from differences in optimization strategy. ScaleHLS and HIDA adopt ping-pong-based dataflow and overlook potential opportunities in more efficient FIFO-based dataflow. POM enhances loop parallelism and resource reuse, but its strategy quickly reaches a performance ceiling as model size grows. Allo depends on user-specified schedules, often leading to suboptimal pipelines and partitions.
In contrast, CODO eliminates both coarse- and fine-grained dataflow violations and prioritizes FIFO implementation, generating high-performance HLS designs.


\textbf{Resource Usage Comparison.} 
As shown in Tables \ref{tab:syn small DNN} and \ref{tab:syn large DNN}, CODO significantly reduces BRAM usage compared to ScaleHLS and HIDA, benefiting from the better resource efficiency of FIFOs over ping-pong buffers. Allo shows 0\% BRAM usage because its aggressive parallelism strategy partitions buffers into very small arrays that Vitis HLS maps to LUTRAM, leading to increased LUT consumption.
CODO delivers the highest performance with similar or lower resource usage. This efficiency stems from two factors. First, compared to ping-pong-based designs such as ScaleHLS, CODO's FIFO-based dataflow minimizes on-chip communication latency and stores only in-flight data, thereby achieving higher performance with reduced resource consumption, as discussed in Section \ref{sec:background-violation}. Second, compared to other FIFO-based approaches such as Allo, CODO eliminates more dataflow violations through the co-optimization of correctness, communication, and parallelism, resulting in improved performance. In addition, the parallelism downscaling step removes redundant optimizations on non-critical loops, balancing the dataflow while conserving computation and memory resources.

\begin{table}
    \renewcommand{\arraystretch}{1.0}
\centering
\caption{On-board Verification of DNN models. 
}
\vspace{-0.3cm}
\label{tab:on-board}
\resizebox{0.99\columnwidth}{!}{
\begin{tabular}{ccccccc}
\toprule[1pt]
\multirow{2}{*}{\begin{tabular}[c]{@{}c@{}}\textbf{Application} \\ \textbf{(input size)}\end{tabular}} &  \multirow{2}{*}{\begin{tabular}[c]{@{}c@{}}\textbf{Frame-} \\ \textbf{work} \end{tabular}} & \multirow{2}{*}{\begin{tabular}[c]{@{}c@{}}\textbf{Overall} \\ \textbf{Speedup} \end{tabular}}   &  \multirow{2}{*}{\begin{tabular}[c]{@{}c@{}}\textbf{Comp.} \\ \textbf{Speedup} \end{tabular}} & \multirow{2}{*}{\begin{tabular}[c]{@{}c@{}}\textbf{P(W)} \end{tabular}} & \multirow{2}{*}{\begin{tabular}[c]{@{}c@{}}\textbf{Exec.} \\ \textbf{ Time(s)} \end{tabular}}   &  \multirow{2}{*}{\begin{tabular}[c]{@{}c@{}}\textbf{E(J)} \end{tabular}}    \\ \\ \toprule[1pt]

\multirow{2}{*}{\begin{tabular}[c]{@{}c@{}}\textbf{ResNet-18}\\ \textbf{(3*32*32)}\end{tabular}} 
& Baseline      & 1$\times$   & 1$\times$  &   34.7&1.839 & 63.8  \\ 
&\textbf{CODO{} }  &\textbf{69.2$\times$}    & \textbf{70.9$\times$}  & 30.7 & 0.026 & 0.8  \\  \hline
\multirow{2}{*}{\begin{tabular}[c]{@{}c@{}}\textbf{VGG-16}\\ \textbf{(3*32*32)}\end{tabular}}  
& Baseline            & 1$\times$  & 1$\times$  &  33.9 & 0.646 & 21.9 \\
& \textbf{CODO{} }           & \textbf{51.0$\times$}   & \textbf{55.9$\times$}  &  31.1&0.013& 0.4  \\ \hline
\multirow{2}{*}{\begin{tabular}[c]{@{}c@{}}\textbf{MobileNet}\\ \textbf{(3*32*32)}\end{tabular}}  
& Baseline             &  1$\times$  & 1$\times$  &  34.5 & 0.041& 1.4  \\
& \textbf{CODO{} }            & \textbf{9.6$\times$}   & \textbf{9.8$\times$}  &   31.0&0.004 & 0.1  \\ 
 \hline
\multirow{3}{*}{\begin{tabular}[c]{@{}c@{}}\textbf{ResNet-18}\\ \textbf{(3*224*224)}\end{tabular}}           
& Baseline            &  1$\times$  &  1$\times$ & 31.8 &7.364& 234.0  \\
& HIDA                 & 23.9$\times$   &  29.1$\times$ &    32.8 &0.308& 10.1  \\
& \textbf{CODO{} }          & \textbf{100.6$\times$}  & \textbf{101.6$\times$}  &    31.1 &0.073& 2.5    \\ \hline
\multirow{3}{*}{\begin{tabular}[c]{@{}c@{}}\textbf{VGG-16}\\ \textbf{(3*224*224)}\end{tabular}}  
& Baseline            &  1$\times$ & 1$\times$  & 34.2 &15.825 & 541.2\\
& HIDA                  & 14.3$\times$   & 82.7$\times$ &  34.2   &1.108& 37.9  \\
& \textbf{CODO{} }          & \textbf{127.5$\times$}   &  \textbf{138.5$\times$}   &  31.2 &0.124& 3.9\\ \hline
\multirow{3}{*}{\begin{tabular}[c]{@{}c@{}}\textbf{MobileNet}\\ \textbf{(3*224*224)}\end{tabular}} 
& Baseline            &  1$\times$   &1$\times$ & 34.7  &1.689& 58.6   \\
& HIDA                & 10.0$\times$    &21.8$\times$ & 31.4 &0.169& 5.3  \\
& \textbf{CODO{} }           & \textbf{43.8$\times$}   & \textbf{44.2$\times$}  &     33.3  &0.039& 1.3     \\ \hline
\multirow{3}{*}{\begin{tabular}[c]{@{}c@{}}\textbf{ZFNet}\\ \textbf{(3*224*224)}\end{tabular}}   
& Baseline            &  1$\times$  & 1$\times$ &   30.8  &3.692& 113.7   \\ 
& HIDA                & 6.2$\times$    & 61.5$\times$  & 32.2&0.593 & 19.1  \\
& \textbf{CODO{} }            &  \textbf{110.7$\times$}  &  \textbf{130.9$\times$}   &  29.9&0.034& 1.0  \\ 
\bottomrule[1pt]
\end{tabular}
}
\vspace{-0.3cm}
\end{table}


\begin{table*}[ht]
\vspace{-0.2cm}
\renewcommand{\arraystretch}{1.0}
\caption{On-board comparison on GPT-2 model. TTFT measures the time to first token, the lower the better. Speed measures the decoding speed in token/s, the higher the better.}
\label{tab:gpt2}
\centering
\resizebox{0.99\textwidth}{!}{
\begin{tabular}{ccccccccccccc}
\toprule[1pt]
\multirow{3}{*}{\begin{tabular}[c]{@{}c@{}}\textbf{[Input Len:}\\ \textbf{Output Len]}\end{tabular}} & \multicolumn{3}{c}{\textbf{DFX}}    & \multicolumn{3}{c}{\textbf{Allo}}   & \multicolumn{3}{c}{\textbf{StreamTensor}} & \multicolumn{3}{c}{\textbf{CODO}}   \\
\cmidrule(lr){2-4}
\cmidrule(lr){5-7}
\cmidrule(lr){8-10}
\cmidrule(lr){11-13}
                                                                                       & \textbf{Latency} & \textbf{TTFT} & \textbf{Speed}     & \textbf{Latency} & \textbf{TTFT} & \textbf{Speed}     & \textbf{Latency}   & \textbf{TTFT}   & \textbf{Speed}       & \textbf{Latency} & \textbf{TTFT} & \textbf{Speed}     \\
                                                                                       & \textbf{(ms)}    & \textbf{(ms)} & \textbf{(token/s)} & \textbf{(ms)}    & \textbf{(ms)} & \textbf{(token/s)} & \textbf{(ms)}      & \textbf{(ms)}   & \textbf{(token/s)}   & \textbf{(ms)}    & \textbf{(ms)} & \textbf{(token/s)} \\
\midrule
\textbf{[32:32]}                                                                       &350.00&177.20&185.19&238.32&81.50&204.05&194.99&34.59&199.51&\textbf{158.64}&\textbf{20.40}&\textbf{231.48}    \\
\textbf{[64:64]}                                                                       &694.70& 349.10&185.19&476.64& 162.99&204.05&358.24&61.27&215.51&\textbf{313.44}&\textbf{32.64}&\textbf{231.48}      \\
\textbf{[128:128]}                                                                     &1384.00&692.80&185.19&953.28&325.98&204.05&696.65&125.35&224.05&\textbf{663.36}&\textbf{110.40}&\textbf{231.48}      \\
\bottomrule[1pt]
\end{tabular}
}
\end{table*}

\begin{figure*}[htbp]
    \centering
    \begin{minipage}{0.65\linewidth}
        \centering
        \includegraphics[width=\textwidth]{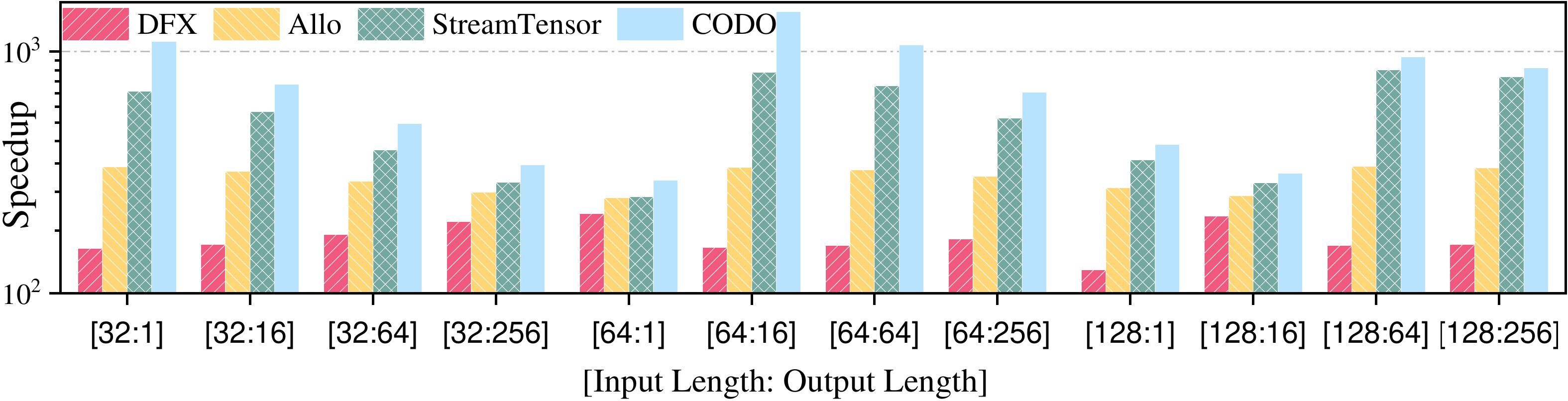}
    \end{minipage}
    \hfill
    \begin{minipage}{0.33\linewidth}
        \centering
        \renewcommand{\arraystretch}{1.0}
        \resizebox{\textwidth}{!}{
        \fontsize{7}{9}\selectfont 
        \begin{tabular}{@{}c@{\hspace{0.2em}}c@{\hspace{0.2em}}c@{\hspace{0.2em}}c@{\hspace{0.2em}}c@{}}
            \toprule
            \textbf{Framework} & 
            \textbf{Platform} & 
            \textbf{HBM} &
            \textbf{Freq.} &
            \textbf{Quant.} \\
            \midrule
            \textbf{DFX} & U280 & 8GB & 200MHz & FP16 \\
            \textbf{Allo} & U280 & 8GB & 250MHz & W4A8 \\
            \textbf{StreamTensor} & U55C & 16GB & 250MHz & W4A8 \\
            \textbf{CODO} & U280 & 8GB & 300MHz & W4A8\\
            \bottomrule
        \end{tabular}%
        }
    \end{minipage}
    \caption{\textbf{Left:} Latency speedup over baseline on the GPT-2 model. \textbf{Right:} Experiment setup of evaluated platforms.}
    \label{fig:gpt2}
    \vspace{-0.3cm}
\end{figure*}

\begin{table}[t]
\centering
\renewcommand{\arraystretch}{1.3}
\caption{Five configurations of optimization methods.}
\resizebox{\columnwidth}{!}{%
\begin{tabular}{lccccc}
\toprule
\textbf{Optimization} & \textbf{Opt1} & \textbf{Opt2} & \textbf{Opt3} & \textbf{Opt4} & \textbf{Opt5} \\
\midrule

\textbf{Coarse-grained Violation Elimination }
& {\color{red}\ding{55}}
& {\color{ForestGreen}\ding{51}}
& {\color{ForestGreen}\ding{51}}
& {\color{ForestGreen}\ding{51}} 
& {\color{ForestGreen}\ding{51}} \\

\textbf{Fine-grained Violation Elimination}  
& {\color{ForestGreen}\ding{51}} 
& {\color{red}\ding{55}}
& {\color{red}\ding{55}} 
& {\color{ForestGreen}\ding{51}} 
& {\color{ForestGreen}\ding{51}} \\

\textbf{Efficient Data Communication}
& {\color{red}\ding{55}} 
& {\color{red}\ding{55}} 
& {\color{ForestGreen}\ding{51}} 
& {\color{ForestGreen}\ding{51}} 
& {\color{ForestGreen}\ding{51}} \\

\textbf{Automated Dataflow Scheduling}
& {\color{red}\ding{55}} 
& {\color{red}\ding{55}} 
& {\color{red}\ding{55}} 
& {\color{red}\ding{55}} 
& {\color{ForestGreen}\ding{51}} \\

\bottomrule
\vspace{-0.2cm}
\end{tabular}%
}
\label{tab:codo_opt_levels}
\end{table}

\begin{figure*}
    \centering
    \includegraphics[width=0.99\textwidth]{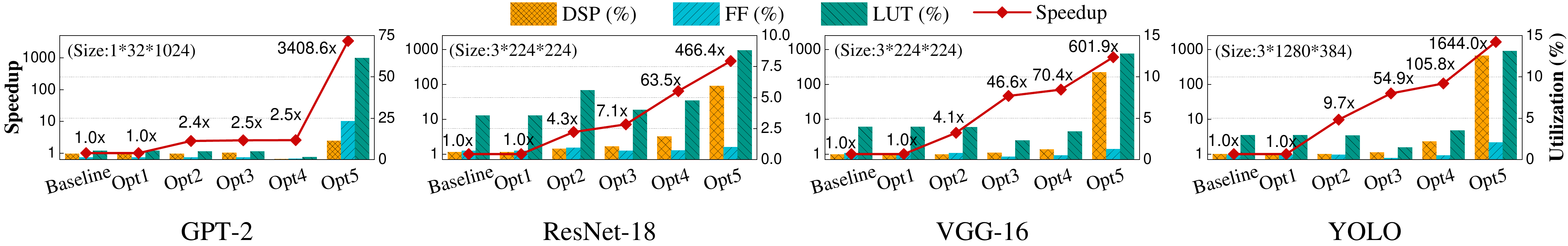}
    \vspace{-0.2cm}
    \caption{Ablation study of different optimization methods.}
    \vspace{-0.3cm}
    \label{fig:abl_methods}
\end{figure*}

\textbf{On-board Verification.} 
We validate CODO's ability to generate deployable dataflow accelerators on an AMD Alveo U280 FPGA. Among the compared frameworks, ScaleHLS fails due to excessive memory usage, Allo encounters deadlocks from fine-grained dataflow violations, and StreamHLS cannot generate valid designs for any workload.
CODO and HIDA successfully generate executable accelerators for all models except YOLO, where inaccurate resource estimation in Vitis HLS prevents implementation.
Table \ref{tab:on-board} shows the on-board evaluation results of DNN models, including runtime speedup, power and energy consumption, and execution time. 
The overall speedup denotes end-to-end performance, including off-chip data communication, whereas comp. speedup focuses exclusively on computation kernels.
CODO's higher performance also enables it to achieve the lowest energy consumption, with $ 77.1 \times$ and $ 9.2 \times$ energy efficiency on average compared to the baseline and HIDA, respectively.

\subsection{Evaluation on the GPT-2 Model}\label{On-board Verification}
To demonstrate CODO's applicability to LLM workloads, we evaluate performance on GPT-2 Medium \cite{gpt2medium_hf}.  Allo and DFX provide manually optimized implementations of GPT-2, and StreamTensor automatically generates optimized GPT-2 accelerators. The remaining compilers do not support transformer models and are therefore excluded from comparison.

The evaluation setup and latency speedups over Vitis HLS-optimized baselines across different input/output sequence lengths are summarized in Fig. \ref{fig:gpt2}. 
Table \ref{tab:gpt2} provides a detailed comparison, showing \texttt{TTFT} for prefill performance and \texttt{Speed} for decoding performance.

Compared to manually optimized designs, CODO provides substantial gains. DFX and Allo require hand-crafted implementations and manual coordination of kernels and data transfers, which is time-consuming, error-prone, and often leads to suboptimal dataflow. In contrast, CODO's automatic DSE quickly generates FIFO-based accelerators, resulting in 3.54$\times$ and 2.03$\times$ speedups over DFX and Allo, respectively.

CODO also outperforms StreamTensor, achieving a 1.23$\times$ speedup, even though StreamTensor runs on a more advanced U55C FPGA.
Although StreamTensor supports automated DSE, it reverts to a ping-pong execution strategy whenever fine-grained violations are detected, which ultimately limits overall performance. In contrast, CODO eliminates these violations as much as possible and applies communication optimizations to ensure high data transfer efficiency. As a result, CODO surpasses StreamTensor even though StreamTensor executes on an FPGA with larger HBM capacity.

\begin{figure}
    \centering
    \includegraphics[width=0.49\textwidth]{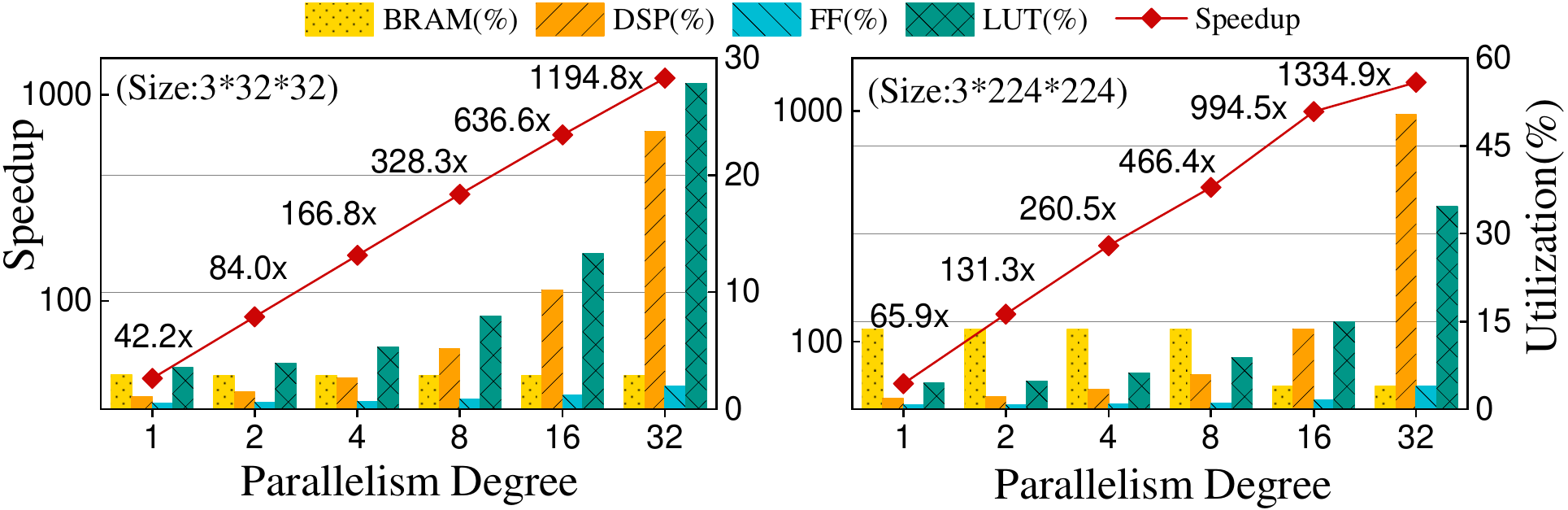}
    \vspace{-0.2cm}
    \caption{Latency speedup and resource usage of ResNet-18 under different degrees of parallelism.}
    \vspace{-0.4cm}
    \label{fig:parallelism ablation}
\end{figure}

\begin{figure}
    \centering
    \includegraphics[width=0.5\textwidth]{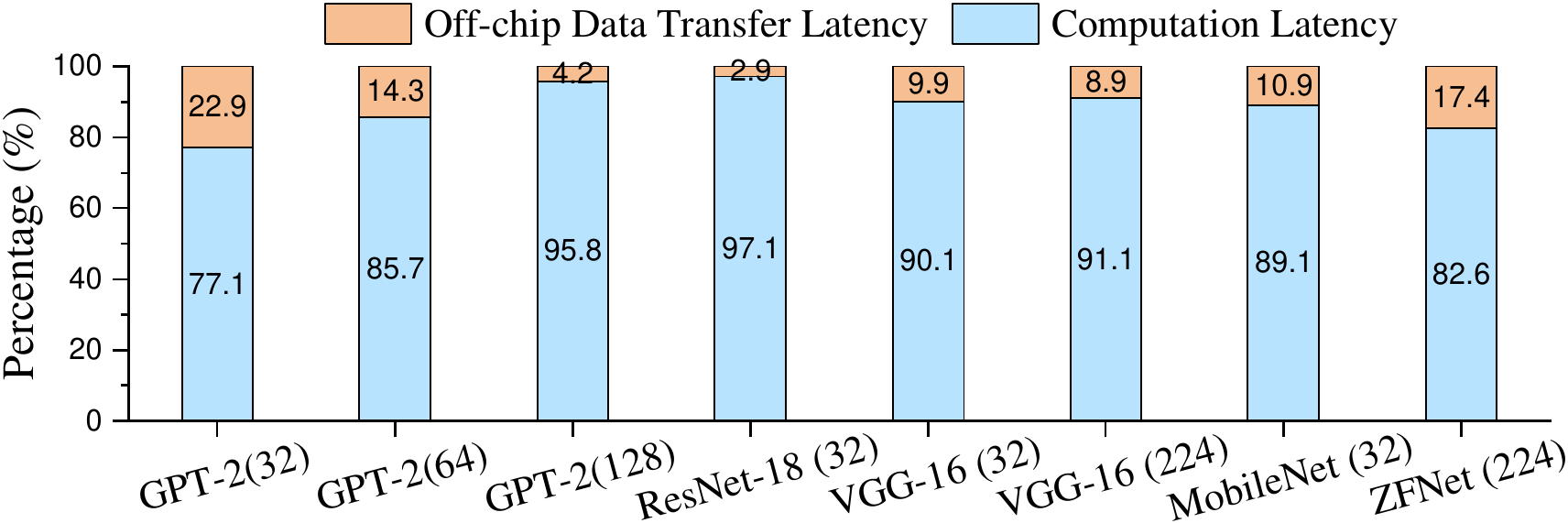}
    \vspace{-0.2cm}
    \caption{On-board execution time breakdown.}
    \vspace{-0.4cm}
    \label{fig:breakdown}
\end{figure}

\begin{table}
    \vspace{-0.3cm}
    \renewcommand{\arraystretch}{1.1}
    \centering
    \caption{Percentage of FIFO usage.}
    \vspace{-0.25cm}
    \label{tab:channel usage}
    \resizebox{\columnwidth}{!}{%
    \begin{tabular}{@{}ccccccc@{}}
    \toprule[1pt]
    \multirow{2}{*}{\begin{tabular}[c]{@{}c@{}}\textbf{Application} \end{tabular}} &  
    \multirow{2}{*}{\begin{tabular}[c]{@{}c@{}}\textbf{Gesummv}\end{tabular}} & 
    \multirow{2}{*}{\begin{tabular}[c]{@{}c@{}} \textbf{Residual}\\ \textbf{Block} \end{tabular}} &
    \multirow{2}{*}{\begin{tabular}[c]{@{}c@{}} \textbf{Multi-Head} \\ \textbf{Attention} \end{tabular}}  & 
    \multirow{2}{*}{\begin{tabular}[c]{@{}c@{}} \textbf{MobileNet} \end{tabular}} & 
    \multirow{2}{*}{\begin{tabular}[c]{@{}c@{}} \textbf{ResNet-18} \end{tabular}} &
    \multirow{2}{*}{\begin{tabular}[c]{@{}c@{}} \textbf{GPT-2} \end{tabular}}\\ \\
    \midrule
    \textbf{\textbf{Percentage}} & 100\% & 100\% & 84\% & 100\% & 100\%& 89\%\\
    \bottomrule[1pt]
    \end{tabular}%
    }
    \vspace{-0.3cm}
\end{table}

\subsection{Ablation Study}
\textbf{Optimization Method Ablation.}
To understand the impact of different optimization methods in CODO, we conduct an ablation study across five configurations (\texttt{Opt1}-\texttt{Opt5}), as defined in Table \ref{tab:codo_opt_levels}. The performance speedups and resource utilization from synthesis results are detailed in Fig. \ref{fig:abl_methods}. Starting with \texttt{Opt1}, we observe that enabling fine-grained optimizations in isolation yields negligible speedup. This is because unresolved coarse-grained violations invalidate dataflow optimization, leading to sequential execution. In contrast, \texttt{Opt2} resolves these violations and enables basic ping-pong buffer-based dataflow execution, achieving initial performance gains ranging from $2.5\times$ to $9.7\times$.
Next, \texttt{Opt3} enables efficient data communication. For models with high data reuse potential, such as ResNet-18 and YOLO, the generation of line and window buffers significantly improves on-chip data reuse and communication efficiency, delivering higher speedups compared to \texttt{Opt2}. Building on this, \texttt{Opt4} addresses fine-grained violations to enable efficient FIFO-based dataflow, boosting performance up to $105.8\times$
Finally, \texttt{Opt5} delivers the highest performance improvements by leveraging resource-aware parallelism exploration and inter-task optimization. Notably, for computation- and communication-intensive workloads like GPT-2, applying \texttt{Opt2}-\texttt{Opt4} yields limited gains due to the extremely imbalanced dataflow, which severely hinders overall performance.  \texttt{Opt5} addresses this by enforcing resource-aware parallelism exploration and inter-task optimization, enabling a high percentage of efficient FIFO implementations with a balanced dataflow. This demonstrates that the superior performance of CODO results from the joint co-optimization of correctness, communication, and parallelism.

\textbf{Resource-Performance Trade-off Evaluation.} To evaluate CODO's ability to generate efficient designs under various resource budgets, we conducted a resource-performance trade-off experiment by adjusting the parallelism degree to simulate different resource budgets, as shown in Fig. \ref{fig:parallelism ablation}. The results show that performance speedup increases nearly linearly with higher parallelism degrees, accompanied by a steady rise in DSP utilization. This indicates that even on resource-constrained FPGA boards, CODO can generate efficient dataflow accelerators by appropriately tuning parallelism.

\textbf{On-board Execution Time Breakdown.} 
Figure \ref{fig:breakdown} shows a detailed breakdown of execution time for GPT-2 with different prefill lengths and DNNs with different input sizes. Overall, data transfer time remains low since CODO effectively utilizes HBM bandwidth. For GPT-2, the data transfer portion is relatively high at short prefill lengths but drops quickly as the  sequence length grows. This is because computation in self-attention increases much faster than data movement, causing computation latency to dominate at larger prefill lengths. With efficient FIFO implementation and communication optimizations, CODO consistently delivers satisfying performance gains across different input lengths.



\textbf{FIFO Percentage Quantification.}\label{Comparison of Dominant buffer Types}
To quantitatively evaluate the effectiveness of our approach, Table \ref{tab:channel usage} reports the proportion of FIFOs used across benchmarks. Except for a few cases in attention and GPT-2, where detected optimization strategy conflicts trigger a fallback to ping-pong buffers, all other tasks achieve a 100\% FIFO implementation. This demonstrates the strong scalability and effectiveness of CODO's dataflow violation elimination.

\vspace{-0.1cm}
\section{Conclusion}
\vspace{-0.1cm}
In this paper, we propose CODO, an automated compiler that detects and resolves dataflow violations in the input DNN models, and generates feasible and efficient dataflow accelerators on FPGAs. CODO provides both on- and off-chip optimizations to improve data communication efficiency. An resource-aware automated scheduling method is equipped to generate high-performance dataflow accelerators rapidly.

\section*{Acknowledgements}
This work was supported by the National Natural Science Foundation of China under Grant 62472273 and Grant 62232015, and the National Key R\&D Program of China under Grant 2022YFB4501400.
\clearpage
\section*{Artifact Appendix}
The whole compilation stack of CODO will be made available as an open-source project on GitHub soon. Users can download and build the project from the source code. Additionally, a docker image containing a pre-built CODO stack is accessible on Docker Hub, facilitating the reproduction of the experimental results in the paper. To enable fast and convenient compilation and execution, a collection of scripts is provided. Furthermore, a unified script is available for automating the entire workflow of the experiment.
\subsection{Artifact check-list}
\label{sec:link}
{\small
\begin{itemize}
  \item {\bf Compilation: } CMake is used to compile the whole project. Our experiments were conducted using CMake version 3.20.3. While versions later than 3.14 are expected to be compatible, we recommend using version 3.20.3 or later for consistency.
  \item {\bf Run-time environment: } Ubuntu 20.04.6 LTS is compatible for the experiments. Other Linux distributions may also work but have not been tested.
  \item {\bf Metrics: } Latency, speedup, compilation time, and resource (BRAM, DSP, FF, and LUT) usage. 
  \item {\bf Output: }  The experimental results will be displayed in the command line output and will also be saved to the corresponding CSV file.
  \item {\bf Experiments: } The key results in the paper are reproduced, including results in TABLE II, III, and IV and Fig. 11. A total of 82 experiments are conducted to reproduce these results.
  \item {\bf How much disk space required (approximately)?: } About 20GB for the Docker image and another 120GB for the Vitis\_HLS/Vivado tools.
  \item {\bf How much time is needed to prepare workflow (approximately)?: } About 10 minutes to download the docker image.
  \item {\bf How much time is needed to complete experiments (approximately)?: } About 20 hours to complete all synthesis experiments.
  \item {\bf Publicly available?: } Yes, the source code of CODO will be released on \href{https://github.com/sjtu-zhao-lab/codo-artifact}{GitHub}, \href{https://hub.docker.com/r/xzz11/codo_ae_image}{Docker Hub}, and \href{https://doi.org/10.5281/zenodo.19425920}{Zenodo}.
\end{itemize}
}

\subsection{Description}

\subsubsection{How to access} 
A docker image containing a pre-built CODO stack is accessible on Docker Hub for users to try out CODO quickly. The links to the repositories are listed in Section \ref{sec:link} in the appendix. The repository is also publicly archived on \href{https://doi.org/10.5281/zenodo.19425920}{Zenodo}.


\subsubsection{Software dependencies} 
Xilinx Vitis HLS 2023.2 and Xilinx Vivado 2023.2 are required in the environment. Please refer to the \href{https://www.xilinx.com/support/download/index.html/content/xilinx/en/downloadNav/vitis/archive-vitis.html}{Xilinx Vitis archive page} for installation instructions.
Note that versions earlier than 2023.2 may yield inconsistent experimental results with results in the paper. The experiments are conducted on a Ubuntu 20.04.6 LTS system. It is highly likely that other Ubuntu versions such as 18.04 and 22.04 are compatible with CODO.



\subsection{Installation}
The docker image can be downloaded from the Docker Hub repository by the instructions below: 
\begin{lstlisting}[language=bash]
  $ docker pull xzz11/codo_ae_image:v1
\end{lstlisting}
When building a new docker container, the directory of Vitis HLS and Vivado need to be mounted.  You can verify your Vitis directory with the following instructions:
\begin{lstlisting}[language=bash]
  $ ls $(YOUR_VITIS_DIR)
    DocNav  Downloads  Model_Composer  
    Vitis  Vitis_HLS  Vivado  xic
\end{lstlisting}
Then the docker container can be built from the provided docker image: 
\begin{lstlisting}[language=bash]
  $ docker run -it -v $(YOUR_VITIS_DIR):$
    (YOUR_VITIS_DIR) -e LC_ALL=en_US.UTF-8 
    -e LANG=en_US.UTF-8 
    xzz11/codo_ae_image:v1 /bin/bash
\end{lstlisting}

\subsection{Experiment workflow}
First, build the compilation environment:
\begin{lstlisting}[language=bash]
$ bash ./compail.sh
\end{lstlisting}

\noindent This step takes approximately 2 minute.\\

A unified script \texttt{./run\_ae.sh} is provided to automate the entire workflow. Alternatively, users can run all synthesis experiments under the \texttt{experiments} directory and extract results as follows:

\textbf{Reproducing all synthesis experiments:}
\begin{lstlisting}[language=bash]
$ cd experiments/
$ bash ./run_all.sh
$ bash ./merge_results.sh
\end{lstlisting}
The extracted results are stored in \texttt{all\_result.csv}.\\

Alternatively, users can reproduce specific figures or tables individually. \\

\textbf{Reproducing Fig. 11:}
\begin{lstlisting}[language=bash]
$ cd experiments/fig-11/
$ bash run_hls.sh
$ python3 extract_rpt_metrics.py
\end{lstlisting}
The execution time is approximately 5 hours. Results are stored in \texttt{result.csv}.\\

\textbf{Reproducing Table II:}
\begin{lstlisting}[language=bash]
$ cd experiments/table-2/
$ bash run_codo.sh
$ bash run_syn.sh
$ python3 batch_extract_rpt_metrics.py
\end{lstlisting}
The execution time is approximately 2 hours. Results are stored in \texttt{result.csv}.\\

\textbf{Reproducing Tables III and IV:}
\begin{lstlisting}[language=bash]
$ cd experiments/table-3\_and\_table-4/
$ bash run_ae.sh
$ bash run_hls.sh
$ python3 batch_extract_rpt_metrics.py
\end{lstlisting}
The synthesis time is approximately 512 minutes. Results are stored in \texttt{result.csv}.\\

\textbf{On-board evaluation results (Table V, Table VI, and Figure 9):} 

The on-board experiments were not conducted for AE purposes, as generating all bitstreams for the evaluated models requires over two weeks and a properly configured U280 environment. Nevertheless, we provide the host and kernel source code, placement-and-route reports, and prebuilt xclbin files for GPT‑2 (corresponding to Table VI and Figure 9).

\begin{lstlisting}[language=bash]
$ cd <corresponding_folder>
$ ./host.exe kernel.hw.xclbin
\end{lstlisting}

\textbf{Reproducing Fig. 9 (on-board execution):}
\begin{lstlisting}[language=bash]
$ cd <corresponding_folder>
$ ./host.exe kernel.hw.xclbin
\end{lstlisting}

\textbf{Reproducing all synthesis experiments:}
\begin{lstlisting}[language=bash]
$ cd experiments/
$ bash run_all.sh
$ bash extract_rpt_metrics.py
\end{lstlisting}

\clearpage
\bibliographystyle{IEEEtranS}
\bibliography{refs}

\end{document}